\documentclass[10pt,preprint]{aastex} 

\usepackage{lscape}

\newcommand{\ovi}{O$\;${\small\rm VI}\relax}

\newcommand{\civ}{C$\;${\small\rm IV}\relax}
\newcommand{\feii}{Fe$\;${\small\rm II}\relax}
\newcommand{\SiII}{Si$\;${\small\rm II}\relax}
\newcommand{\htwo}{H$_2$}

\newcommand{\halpha}{H$\alpha$}

\newcommand{\HI}{H$\;${\small\rm I}\relax}
\newcommand{\HII}{H$\;${\small\rm II}\relax}
\newcommand{\lyb}{Ly$\beta$}

\newcommand{\wave}[1]{$\lambda$#1\relax}

\newcommand{\msun}{M$_\odot$}
\newcommand{\kms}{km~s$^{-1}$\relax}

\newcommand{\percc}{cm$^{-3}$\relax}
\newcommand{\column}{cm$^{-2}$}

\newcommand{\dg}{$^\circ \,$}
\newcommand{\sk}{Sk--}

\newcommand{\e}[1]{10^{#1}}

\newcommand{\fuse}{{\em FUSE}}
\newcommand{\rosat}{{\em ROSAT}}
\newcommand{\copernicus}{{\em Copernicus}}
\newcommand{\hst}{{\em HST}}
\newcommand{\iue}{{\em IUE}}
\newcommand{\calfuse}{{\tt CALFUSE}}

\begin{document}


\title{The Global Content, Distribution, and Kinematics of 
	Interstellar \ion{O}{6} in the Large Magellanic Cloud}

\author{J. Christopher Howk\altaffilmark{1}, 
  	Kenneth R. Sembach\altaffilmark{1,2}, 
	Blair D. Savage\altaffilmark{3},
	Derck Massa\altaffilmark{4}, 
	Scott D. Friedman\altaffilmark{1},  \&
	Alex W. Fullerton\altaffilmark{1,5}}

\altaffiltext{1}{Department of Physics and Astronomy,
The Johns Hopkins University, Baltimore, MD, 21218; howk@pha.jhu.edu,
scott@pha.jhu.edu, awf@pha.jhu.edu}

\altaffiltext{2}{Current address: Space Telescope Science Institute, 
3700 San Martin Dr., Baltimore, MD 21218; sembach@stsci.edu}

\altaffiltext{3}{Astronomy Department, University of Wisconsin-Madison,
Madison, WI, 53711; savage@astro.wisc.edu}

\altaffiltext{4}{Emergent Information Technologies, Code 681, NASA's 
	GSFC, Greenbelt, MD 20771; derck.massa@gsfc.nasa.gov}


\altaffiltext{5}{Department of Physics and Astronomy,
                 University of Victoria, P.O. Box 3055, Victoria, BC
                 V8W 3P6, Canada}


\begin{abstract}

We present {\em Far Ultraviolet Spectroscopic Explorer} (\fuse)
observations of interstellar \ovi\ absorption towards 12 early-type
stars in the Large Magellanic Cloud (LMC).  The observations have a
velocity resolution of $\la20$ \kms\ (FWHM) and clearly show
\ovi\ 1031.926 \AA\ absorption at LMC velocities towards all 12 stars.  
From these observations we derive column densities of interstellar
\ovi\ in this nearby galaxy; the observed columns are in the range
$\log N(\mbox{\ovi}) = 13.9$ to 14.6, with a mean of 14.37 and a
standard deviation of $\pm38\%$ ($^{+0.14}_{-0.21}$ dex).  The
observations probe several sight lines projected onto known
superbubbles in the LMC, but these show relatively little (if any)
enhancement in \ovi\ column density compared to sight lines towards
relatively quiescent regions of the LMC.  The observed LMC \ovi\
absorption is broad, with Gaussian dispersions $\sigma \approx 30$ to
50 \kms.  This implies temperatures $T\la(2-5)\times10^6$ K,
indicating that much of the broadening is non-thermal because \ovi\
has a very low abundance at such high temperatures. The \ovi\
absorption is typically displaced $\sim-30$ \kms\ from the
corresponding low-ionization absorption associated with the bulk of
the LMC gas.

The general properties of the LMC \ovi\ absorption are very similar to
those of the Milky Way halo.  The average column density of \ovi\ and
the dispersion of the individual measurements about the mean are
identical to those measured for the halo of the Milky Way, even though
the metallicity of the LMC is a factor of $\sim2.5$ lower than the
Milky Way.  The velocity dispersion measured for the LMC material is
also consistent with recent measurements of the Galactic halo.  The
striking similarities in these quantities suggest that much of the LMC
\ovi\ may arise in a vertically-extended distribution similar to the
Galactic halo.  We discuss the measurements in the context of a halo
composed of radiatively cooling hot gas and/or turbulent mixing
layers.  If the observed \ovi\ absorption is tracing a radiatively
cooling galactic fountain flow, the mass flow rate from one side of
the LMC disk is of the order $\dot{M} \sim 1$ \msun\ yr$^{-1}$, with a
mass flux per unit area of the disk $\dot{M} /
\Omega \sim 2\times10^{-2}$
\msun\ yr$^{-1}$ kpc$^{-2}$.

\end{abstract}



\section{Introduction}
\label{sec:intro}

Current theory suggests that the interstellar medium (ISM) in galaxies
is strongly influenced by the input of energy and matter from stars.
This is particularly true for spiral and irregular galaxies which are
typically experiencing on-going star formation.  The effects of the
interactions of (multiple) stars and supernovae with the ISM in spiral
and irregular galaxies is expected to produce pockets of super-heated
gas (e.g., Heiles 1990; Norman \& Ikeuchi 1989; Mac Low \& McCray
1988), and the holes visible in the neutral hydrogen distributions of
galaxies are often interpreted as the signature of such effects (Kim
et al. 1999; Puche et al. 1992; Deul \& den Hartog 1990; Brinks \&
Bajaja 1986).  In regions with high densities of early-type stars, the
energy input from the stars and supernovae can shape interstellar
matter on kiloparsec scales.  For disk galaxies, with larger pressure
gradients in the vertical, $z$, direction than the radial direction,
the super-heated, high pressure gas is thought to flow away from the
disk, feeding matter and energy into the thick disk or halo of the
system (e.g., de Avillez 2000; Norman \& Ikeuchi 1989; Bregman 1980).

The Large Magellanic Cloud (LMC) is the nearest disk galaxy to the
Milky Way.  Its proximity ($d\sim50$ kpc) and low inclination angle
($i\approx30^\circ-40^\circ$) provide a relatively clear view of this
system (see Westerlund 1997), with very little foreground or
line-of-sight confusion.  The LMC has long served as a laboratory for
studies of the ISM.  Excellent maps of the ISM within the LMC are
available for the warm ($\la 10^4$ K) gas, both neutral (Kim et
al. 1998) and ionized (e.g., Smith 1999; Kennicutt et al. 1995;
Davies, Elliott, \& Meaburn 1976), as well as the hot gas (Snowden \&
Petre 1994; Wang et al. 1991).  The ability to resolve individual
stars from each other and the ISM in this system provides a unique
opportunity to determine the energy budget of the largest interstellar
structures within the galaxy (e.g., Oey 1996).

Studies of the hot gaseous content of the LMC are of particular
importance for understanding the energy input into the ISM from stars.
One manner in which to study the hot gas content of a galaxy is
through ultraviolet (UV) absorption line observations of the Li-like
oxygen ion (O$^{+5}$ or \ovi), which has a doublet of transitions at
1031.926 and 1037.617 \AA.  The \ovi\ ion principally arises in hot,
collisionally-ionized gas in galactic environments since the
ionization potential for its production (IP$_{\rm O\, V - O\, VI} =
114$ eV) is high enough to preclude a significant photoionized
component.  Such observations are currently plausible using the {\em
Far Ultraviolet Spectroscopic Explorer} (\fuse), a dedicated
spectroscopic observatory operating in the wavelength range $905 -
1187$ \AA\ at high resolution ($R \equiv \lambda / \Delta
\lambda \approx 15,000-20,000$).  \fuse\ can easily observe stars in
the Galactic halo, stars in the Magellanic Clouds, and distant
extragalactic sources with sufficient signal to noise to produce
reliable \ovi\ column density and line profile measurements.

Here we present \fuse\ spectra of \ovi\ absorption towards 12
early-type stars in the LMC, the properties of which are summarized in
Table \ref{tab:targets}.  In this work we show that the LMC contains a
large amount of highly-ionized gas distributed in a patchy manner
across its face.  The \ovi\ column densities in the LMC along the
directions studied in this work are as high as those observed through
the halo of the Milky Way (Savage et al. 2000).  We show that the hot
material traced through \ovi\ is present at significantly different
velocities than the bulk of the low-ionization material arising in the
disk of the LMC, a phenomenon that was also seen in {\em Hubble Space
Telescope} (\hst) observations of \civ\ absorption (Wakker et
al. 1998).  We will argue that the available evidence suggests much of
the observed LMC absorption arises in an extended halo or corona,
perhaps very similar to that observed in the Milky Way.  These results
are important since the LMC is the only other disk galaxy beyond the
Milky Way that can be studied in this manner.

In \S \ref{sec:data} we discuss the reduction of the \fuse\ data
employed in this work, while \S \ref{sec:analysis} discusses the
methods used in analyzing the data.  We discuss the gross distribution
of \ovi\ column densities in the LMC in \S \ref{sec:columns}, and the
rough kinematic properties of the hot material in \S
\ref{sec:kinematics}.  In \S \ref{sec:discussion} we discuss the
implications of our observations, and we summarize our work in \S
\ref{sec:summary}.

\section{\fuse\ Observations and Data Reduction}
\label{sec:data}

The \fuse\ mission, its planning and in-orbit performance are
discussed by Moos et al. (2000) and Sahnow et al. (2000).  Briefly,
the \fuse\ observatory consists of four co-aligned prime-focus
telescopes and Rowland-circle spectrographs feeding two microchannel
plate (MCP) detectors with helical double delay line anodes.  Two of
the telescope/spectrograph channels have SiC coatings providing
reflectivity over the range $\sim 905 - 1105$ \AA, while the other two
have Al:LiF coatings for sensitivity in the $\sim1000 - 1187$ \AA\
range.  We make use of only the latter, the LiF channels, in this
work.  Light from each mirror passes through an aperture onto a
holographically-ruled spherical grating.  The resulting spectra are
projected onto the MCP detectors, with spectra from one LiF and one
SiC channel imaged in parallel onto each detector.

The data used in this work were all obtained with the source centered
in the $30\arcsec\times30\arcsec$ (LWRS) aperture of the LiF1
spectrograph channel.\footnote{The Sk--67\dg05 data were obtained as
part of in-orbit checkout activities that required multiple exposures
with the object at different locations within the aperture.  The
processing of this observation required special care to ensure that
the individual exposures were shifted and summed properly.  The
analysis of this sight line is discussed by Friedman et al. (2000).}
In many cases, the LiF2 channel was co-aligned well enough with the
LiF1 channel that a second set of spectra covering the 1000--1187 \AA\
wavelength range was obtained.  Total exposure times ranged from 3.6
ksec to 33 ksec for the LMC stars used in this work.  A log of the
observations used for this investigation is given in Table
\ref{tab:log}.
 
The raw datasets consist of tables of time-tagged photon event
locations within each $16384\times1024$ detector
array.\footnote{Bright targets make use of the spectral image mode
wherein the photons arriving at each pixel are simply summed and read
out at the end of the exposure.  This mode does not preserve the
time-tagged nature of the data.}  The individual exposures for a set
of observations were merged into a single spectrum by concatenating
the individual photon event lists.  These combined time-tagged photon
event lists were processed with version 1.8.6 or 1.8.7 of the standard
FUSE calibration pipeline ({\tt CALFUSE}) available at The Johns
Hopkins University [see Oegerle et al. (2000) for a detailed
description of the {\tt CALFUSE} pipeline].  The differences between
these two versions of the pipeline are inconsequential for our
purposes.  The photon lists were screened for valid data with
constraints imposed for earth limb angle avoidance and passage through
the South Atlantic Anomaly.  Corrections for detector backgrounds,
Doppler shifts caused by spacecraft orbital motions, and geometrical
distortions were applied (Sahnow et al. 2000).  No corrections were
made for optical astigmatism aberrations, and no flatfield corrections
are yet applied in the
\fuse\ calibration process.  Given the existence of significant
fixed-pattern noise structure introduced by the \fuse\ detectors, the
latter omission can be particularly important.  We compared the LiF1
and LiF2 spectra whenever possible to determine the significance of
the observed features.

The processed data have a nominal spectral resolution of $\la20$
\kms\ (FWHM), with a relative wavelength dispersion solution accuracy
of $\sim 6$ \kms\ ($1\sigma$).  The zero point of the wavelength scale
for each individual observation is poorly determined.  In a few cases
{\em International Ultraviolet Explorer} ({\em IUE}) or Space
Telescope Imaging Spectrograph (STIS) data allowed us to constrain the
velocity zero point of observations of LMC stars.  These include the
sight line toward Sk--71\dg45 with STIS and Sk--67\dg211 with {\em
IUE}.  We have also investigated the velocity offsets of several stars
not presented in this work and find that a consistent shift of
$\approx -45$ to $-50$ \kms\ from the \calfuse\ wavelength solution
was often appropriate to bring the \fuse\ data into the LSR frame
(with the exception of the Sk--67\dg05 datasets; see Friedman et
al. 2000).  This is consistent with the average shift of $-47\pm5$
\kms\ derived for \fuse\ LMC/SMC observations calibrated with STIS
echelle-mode observations by Danforth et al. (2002).  We note that
this shift includes an implicit correction for the sign error in the
application of the heliocentric velocity correction in all \calfuse\
versions prior to v2.0, though the magnitude of this error is small
for LMC sight lines (of order $\sim5$
\kms).  In those cases where other data are not available, we have 
adopted a $-50$ \kms\ shift to the \fuse\ velocity scale.

\begin{figure}
\epsscale{1.1}
\plotone{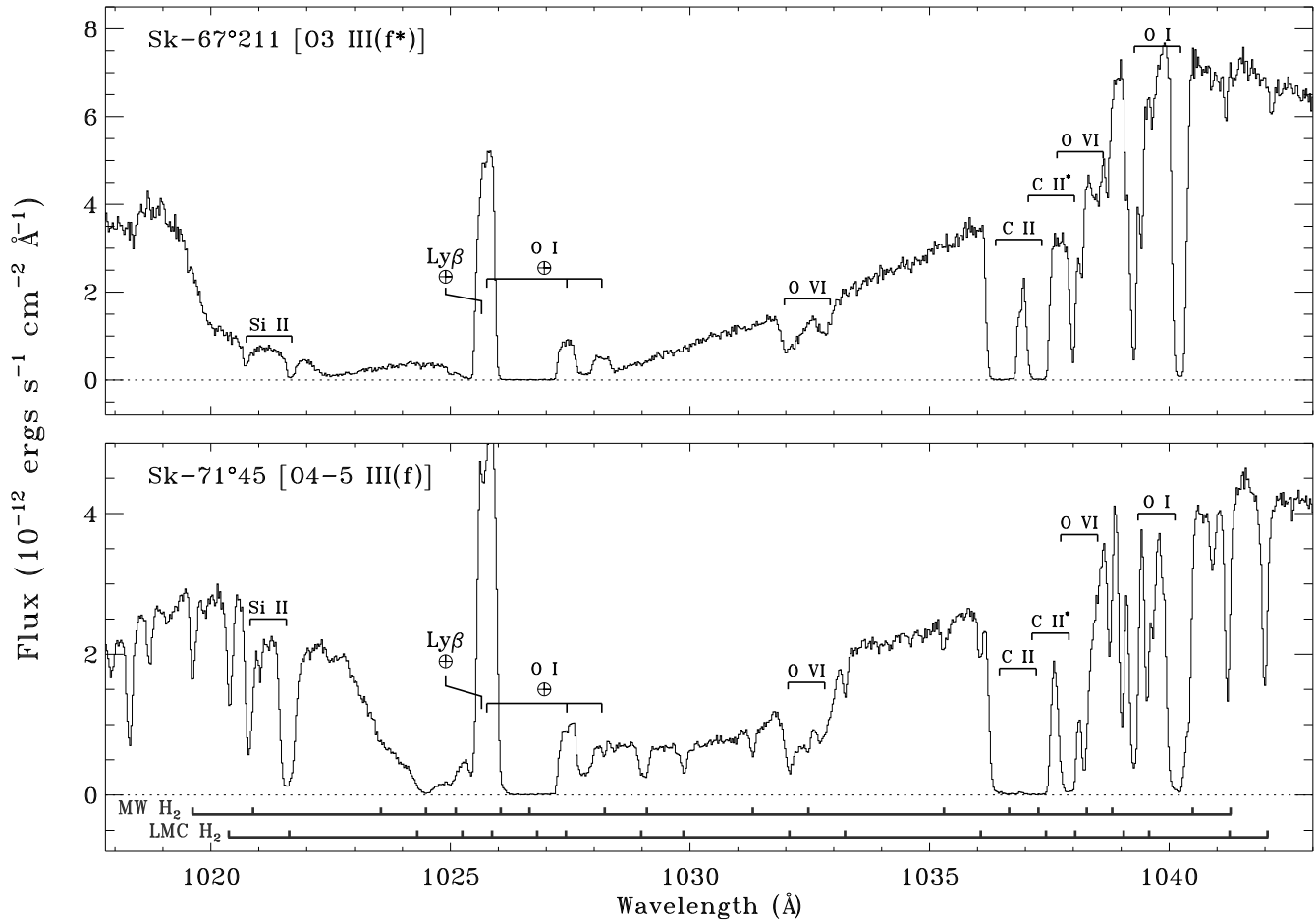}
\caption{Sample \fuse\ spectra in the spectral region centered on 
1030 \AA\ for the stars Sk--67\dg211 ({\em top}) and Sk-71\dg45 ({\em
bottom}).  The data shown in this figure have been rebinned to
$\approx27$ m\AA, or $\sim7.8$ km s$^{-1}$, per pixel for display
purposes.  Several ionic species seen in the ISM of both the Milky Way
and LMC are marked, with the tick separation approximately
corresponding to the velocity separation of the low-ion absorption
along each sight line.  Airglow lines are also indicated.  The sight
line towards Sk--71\dg45 shows a rich H$_2$ spectrum in both the Milky
Way and the LMC.  The expected positions of H$_2$ transitions (for $J
\le 4$ only) are marked below this spectrum.  The sight line towards
Sk--67\dg211 shows only very weak absorption from molecular hydrogen
in both the Milky Way and LMC.  The (6-0) P(3) and R(4) transitions of
H$_2$ at rest wavelengths of 1031.19 and 1032.35 \AA, respectively,
can contaminate the \ion{O}{6} \wave{1031.926} absorption from the
halo of the Milky Way.
\label{fig:fullspec}}
\end{figure}

Figure \ref{fig:fullspec} shows portions of the final \fuse\ spectra
of two stars near the \ovi\ doublet: Sk--67\dg211 and Sk--71\dg45.
These spectra represent only data from the LiF1 channel for each star.
Several prominent ionic absorption lines are marked, with ticks
showing the location of the Milky Way and LMC absorption components.
The positions of the Milky Way and LMC ticks are determined by the
velocity of the \SiII\ line at 1020.699 \AA.  We have not marked the
position of interstellar \lyb\ absorption.  Telluric emission lines
from \lyb\ and \ion{O}{1} can be seen to varying degrees in the data.

These two sight lines represent two extremes of molecular hydrogen
absorption for our sample of stars.  The sight line towards
Sk--71\dg45 shows a rich spectrum of \htwo\ absorption from both Milky
Way and LMC gas.  The expected positions of the \htwo\ lines (for
$\lambda \ga 1018$ \AA\ and $J\le 4$ only) in both the Galaxy and LMC
towards Sk--71\dg45 are marked under the spectrum of that star.  The
sight line towards Sk--67\dg211, however, shows only very weak \htwo\
absorption (e.g., the Lyman (5-0) R(3) transition at 1041.16 \AA).  We
discuss the importance of \htwo\ contamination of the interstellar
\ovi\ profiles in \S \ref{subsec:htwo}.  The properties of the 
\htwo\ absorption within the LMC are discussed by Tumlinson et
al. (2001).

It should be noted from Figure \ref{fig:fullspec} that the absorption
from the weaker \ovi\ transition at 1037.617 \AA\ is almost always
heavily confused by absorption from \ion{C}{2}$^*$ and \htwo\ when
studying the LMC.  This is the case even for Sk--67\dg211, which
exhibits only very weak \htwo\ absorption.

\section{Analysis of Interstellar \ovi\ Absorption}
\label{sec:analysis}

This section discusses the details of our analysis of the interstellar
\ovi\ absorption.  Of particular importance is the estimation of the 
stellar continuum, which is determined by the specifics of the
out-flowing stellar wind for the stars in the sample (see \S
\ref{subsec:continua}).  The estimation of the strength of \htwo\
contamination at the velocities of the interstellar \ovi\ is more
important for absorption at Milky Way velocities than for absorption
at the velocities of the LMC, but in some cases it can be important
for the latter as well (\S \ref{subsec:htwo}).  Finally we discuss the
details of our measurements of equivalent widths and column densities
(\S \ref{subsec:measurements}).

\subsection{Estimation of the Stellar Continuum}
\label{subsec:continua}

The single greatest uncertainty in an analysis of the interstellar
\ovi\ absorption from the LMC is the determination of the stellar 
continuum in the regions immediately surrounding the interstellar
features.  As discussed by Massa et al. (2002), the shape of the local
stellar continuum in the \ovi\ spectral region of early-type stars is
dominated by resonance absorption (and scattering) of the stellar
continuum emission by \ovi\ ions in the out-flowing stellar wind.  The
principle source of \ovi\ within the stellar winds of the O stars is
likely ionization by X-rays produced in strong shocks propagating
through the wind (Puls, Owocki, \& Fullerton 1993).  The radiative
transfer of photons from the photosphere through the expanding wind
yields a characteristic P Cygni-like profile (see Lamers \& Cassinelli
1999).  Because of the stochastic nature of the shock formation
process, the shape of the \ovi\ P Cygni profiles can be significantly
different for individual OB and WR stars and can even be temporally
variable (e.g., Lehner et al. 2001).

Our approach to studying the interstellar \ovi\ in this first study of
the LMC is to use only stars whose stellar continua are
straightforward to approximate.  Furthermore, we have compared our
fits of the stellar continua with the Sobolev Exact Integration (SEI;
see Massa, Prinja, \& Fullerton 1995) modelling of the radiative
transfer of photospheric photons through the out-flowing stellar wind
from Massa et al. (2002).  We have insured that there are no
significant systematic inconsistencies between the the stellar
continua determined when viewing the stellar wind on a relatively
local scale (i.e., within a few hundred \kms\ of the ISM absorption)
and the information derived from self-consistent radiative transfer
models of the entire stellar wind profiles (with the provisos
discussed by Massa et al. 2002).

We have limited our sample to WR stars and O stars with spectral types
O7 and earlier.  The use of only the earliest spectral types implies
that we are using stars with large mass loss rates (and often large
terminal velocities as well), which tend to show more fully developed
(i.e., simpler) \ovi\ P Cygni profiles and are less susceptible to
difficulties caused by wind variations (see discussion below).  To
this end we have examined the existing \fuse\ PI team spectra of all
stars meeting these spectral class requirements.  From this sample we
have chosen to study 12 sight lines: 11 new sight lines and 1
(\sk67\dg05) studied by Friedman et al. (2000).  The properties of
these 12 stars are summarized in Table \ref{tab:targets}.

\begin{figure}
\epsscale{0.9}
\plotone{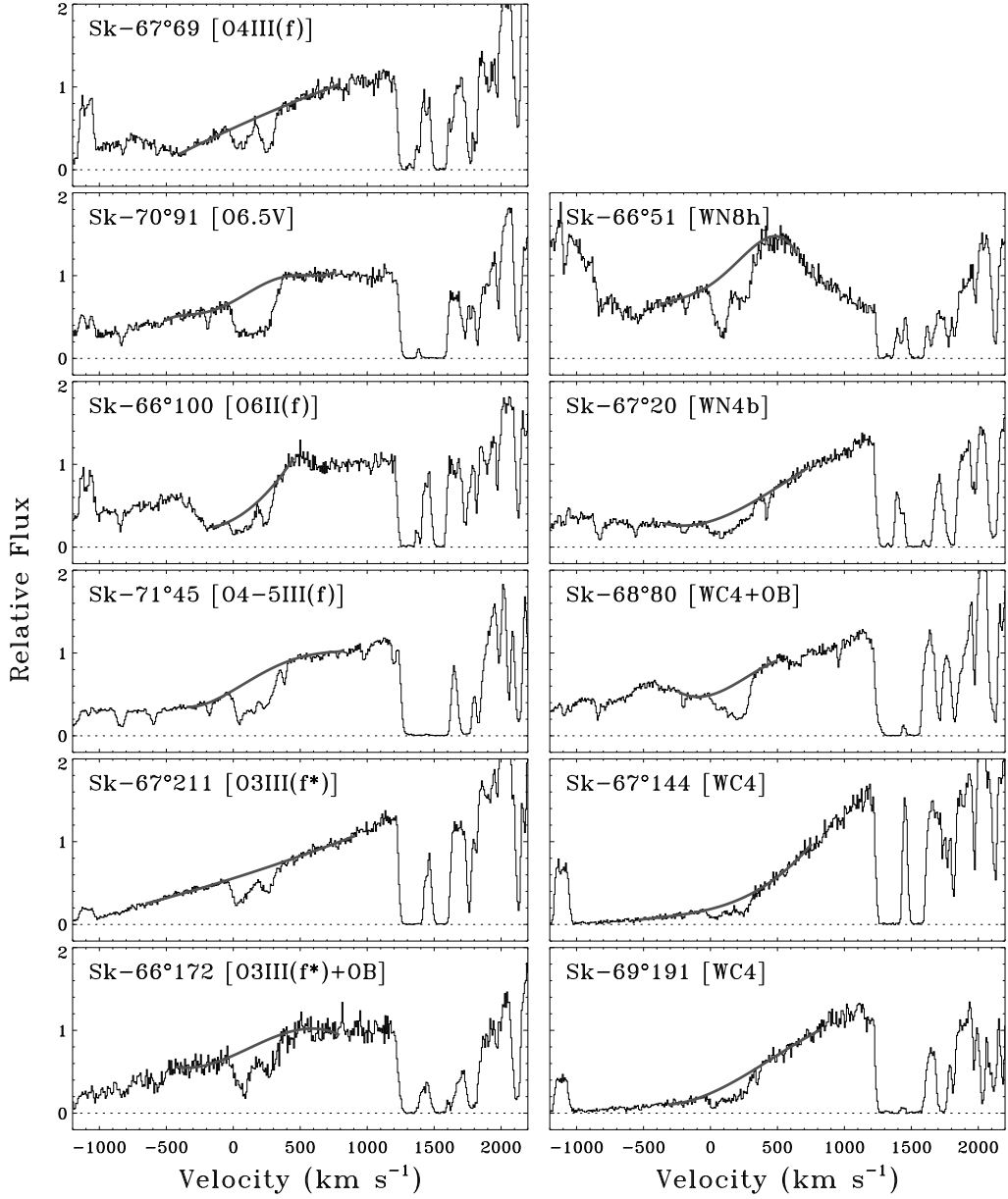}
\caption{Sample fits to the stellar continua in the region of  
interstellar \ion{O}{6} $\lambda1031.926$ absorption for the LMC O
stars ({\em left}) and WR stars ({\em right}) studied in this work.
The velocity scale is (approximately) relative to the LSR, and the
data have been binned to $\approx27$ m\AA, or $\sim7.8$ km s$^{-1}$,
per pixel for display purposes.
\label{fig:continua}}
\end{figure}

Figure \ref{fig:continua} shows the adopted stellar continua for the O
stars (left column) and WR stars (right column) studied in this work.
For each star we estimated the shape and level of the local stellar
continuum near the interstellar \ovi\ \wave{1031.926} absorption by
fitting low-order ($\le5$) Legendre polynomials to adjacent spectral
regions free from interstellar lines.  This fitting followed the
techniques described by Sembach \& Savage (1992), including the
estimation of the uncertainties in the Legendre polynomial fitting
coefficients.  This allows us to include an estimate for continuum
fitting uncertainties in our equivalent width and column density
measurement error estimates (see below).  In each case we adopted the
lowest-order polynomial that could reasonably approximate the stellar
continuum.  Though we applied an $F$-test to assess the significance
of additional polynomial terms, the final decision on the order of the
fit was subjective.

A few general comments regarding the placement of stellar continua for
early-type stars are necessary.  The separation of the two members of
the \ovi\ doublet is $\sim1650$ \kms.  Thus, for stars whose winds
have terminal velocities, $v_\infty$, in excess of this velocity, the
absorption from the red member of the doublet will overlap that of the
blue member.  From the standpoint of understanding the stellar
continuum in the region of interstellar \ovi, stars with stellar wind
terminal velocities within a few hundred \kms\ of the 1650 \kms\
separation of the \ovi\ doublet are often problematic.  Given that the
terminal velocity is often manifested as a relatively sharp gradient
in the absorption strength of the wind profile, stars with such
terminal velocities can have relatively complex stellar continua in
the region of interest for interstellar studies.

For terminal velocities $1650 \la v_\infty \la 2000$ \kms\ the
absorption edge of the 1031.926 \AA\ transition lies in the deep
interstellar Lyman-$\beta$ trough or the 1037 \AA\ transition overlaps
the 1031.926 \AA\ rest velocity.  In both of these cases, the overlap
with other transitions can make it very difficult to obtain a reliable
SEI fit because the models typically try to simultaneously fit both
lines of the doublet.  With the exception of \sk67\dg05, the terminal
velocities for the O stars in Table \ref{tab:targets} are in excess of
these values.  The terminal velocities for the WR stars are all
outside of this range (though we did not make use of SEI modelling of
these stars).  Our exclusion of lower terminal velocity O stars is not
because of the difficulty of the SEI modelling, but rather because
stars with lower terminal velocities often display more complex
continua in the region of interstellar \ovi\ absorption.  However,
studies of the ISM towards later spectral type stars will need to
consider the difficulties in the SEI modelling when comparing such
models to their data.

Another concern regarding the general shape of the stellar wind
absorption in the region of the \ovi\ ISM lines is the probable
variability of the stellar wind \ovi\ absorption profiles.  Lehner et
al. (2001) have discussed such variations and their effects on
continuum placement for interstellar studies.  Using multiple \fuse\
observations of the \ovi\ spectral region in Galactic and LMC stars,
these authors showed that, even in sparsely sampled time series of two
to three spectra, a large fraction ($\ga60\%$) exhibited some spectral
variability over the time periods probed (days to months).  In general
the variability in the \ovi\ P Cygni profiles happened at velocities
much higher than those of the interstellar absorption towards these
stars.  Furthermore, the scale over which the \ovi\ varies was
typically several hundred \kms, significantly broader than most of the
interstellar absorption lines in the Galaxy.  Lehner et al. (2001)
found that even though the shape of the stellar continuum can vary
significantly with time, the derived \ovi\ column densities and
equivalent widths for Galactic absorption were typically not
significantly different.

The situation for absorption in the LMC is less encouraging.  Because
the full extent of the absorption from the Galactic halo and the LMC
typically extends over $\sim300-400$ \kms, the expectation that
changes in the P Cygni profiles should occur over scales larger than
that of the ISM absorption profile is no longer valid.  Because the
Milky Way absorption in these directions occurs at velocities of a few
hundred \kms\ with respect to the rest frame of the stars, it is also
no longer the case that most of the variation should occur at
velocities well beyond the interstellar absorption.  Indeed, the
column densities and equivalent widths estimated towards the one LMC
star in the study by Lehner et al. were significantly different
between the two times sampled.

At this point we are unable to quantitatively assess the affects of
stellar wind variability on the column densities derived along the
sight lines to the LMC stars discussed in this work (see below).  We
note, however, that our choice of the earliest spectral type stars may
help in this regard.  Hot stars with large mass loss rates are
desirable as background probes since their \ovi\ wind lines are
usually saturated.  Nearly all of the flux observed in a saturated
wind line is due to scattered light from throughout the wind.
Therefore, substantial changes in the line of sight wind optical depth
cannot have much of an impact on the flux profile because the
transmitted flux is a small fraction of the total flux. The stellar
wind profiles of these stars are therefore less susceptible to
variations.  The LMC star studied by Lehner et al. (2001) has a
significantly lower terminal velocity and is of a later spectral type
than the stars studied in this work.  The continuum placement for this
star is much more difficult than for any star in the present sample.
This is important insofar as the column densities and equivalent
widths of interstellar \ovi\ can be accurately determined along the
sight line to that star, irrespective of whether or not any variation
in the wind profile is present.

Given the above discussion of wind variability and its effects on the
derivation of \ovi\ column densities, the reader should bear in mind
that the column densities reported in this work have not been
corrected for any stellar wind absorption features that may reside at
velocities similar to the ISM features.  Future re-observations of
some of these stars are planned and may reveal some such features.
However, the presence of discrete stellar wind absorption features at
interstellar velocities can never be ruled out since one can only
identify their presence if they change.

We should also note that for this work we have chosen to consider
targets having relatively unambiguous separation between the stellar
and interstellar features and well-delineated stellar continua, with
the exception of Sk--67\dg05.  The continuum placement and
uncertainties for this star, which also falls outside the range of
spectral types we have investigated, are discussed in detail by
Friedman et al. (2000).  In general the stellar continua are more
complicated as one looks to later spectral types because the
\ovi\ P Cygni profiles of later type stars tend to be less well developed
than those of the early O stars.  As our understanding of the stellar
winds of the later spectral-type stars improves, we may be able to
make use of the larger sample of \fuse\ observations of late O and B
stars in the LMC for studying interstellar \ovi.

\subsection{Assessing the Contamination from Molecular Hydrogen}
\label{subsec:htwo}

Absorption from molecular hydrogen in both the Milky Way and the LMC
can overlap the \ovi\ absorption along the observed sight lines.
Absorption from the (6-0) P(3) and R(4) transitions of \htwo\ at
1031.19 and 1032.35 \AA, respectively, can contaminate the \ovi\
absorption from the Milky Way halo and in some cases the \ovi\
absorption from the LMC (e.g., see Figure \ref{fig:fullspec}).  These
lines absorb at $-214$ and $+123$ \kms\ relative to the rest frame of
\ovi\ \wave{1031.926}.  Rather than derive the detailed characteristics of
the entire \htwo\ absorption spectrum, we have taken a more pragmatic
approach to estimating the contamination of the interstellar \ovi\
profiles by molecular hydrogen.

To estimate the contribution to the \ovi\ profiles from the (6-0) P(3)
and R(4) lines of molecular hydrogen, we have fit Gaussian profiles to
a number of $J=3$ and 4 transitions of \htwo\ with line strengths,
$\log \lambda f$, very similar to the contaminating 6-0 band
transitions.  The \htwo\ transitions used in this approach are given
in Table \ref{tab:htwo}.  For comparison, the properties of the
contaminating (6-0) P(3) and R(4) lines are also given (in bold).  We
have only used lines which appear in the LiF1 channel.  This helps to
minimize the effects of changes in the breadth and shape of the
instrumental line spread function, which occur when comparing lines in
different channels and detector segments.  Because of blending with
other species and varying line strengths, the transitions available
from Table \ref{tab:htwo} varied for each sight line.

The fitting process yields a peak apparent optical depth, a breadth,
and a central velocity for each of the transitions.  We have scaled
the peak apparent optical depth for each fitted transition by the
$\lambda f$ ratio of the contaminating 6-0 band line to that of the
line fitted.  Typically we were able to measure several transitions
and average the scaled peak optical depths, breadths, and velocities
predicted for the contamination (6-0) P(3) and R(4) transitions.  For
the $J=3$ transitions we often measured both weaker and stronger
transitions than the offending line.  Where possible, we have
fine-tuned the velocities using the Milky Way P(3) and LMC R(4)
components on either side of the interstellar \ovi\ profiles.  A
difficulty with this general approach is that saturation effects can
be significant for unresolved spectral lines.  We have attempted to
circumvent such difficulties by judiciously choosing comparison
transitions with line strengths $\lambda f$ close enough to those 6-0
band transitions that the factor by which the peak optical depths are
scaled is small.  In general we have found that the scaled peak
optical depths for the comparison transitions show a smaller
dispersion than the directly measured values.

\begin{figure}
\epsscale{0.85}
\plotone{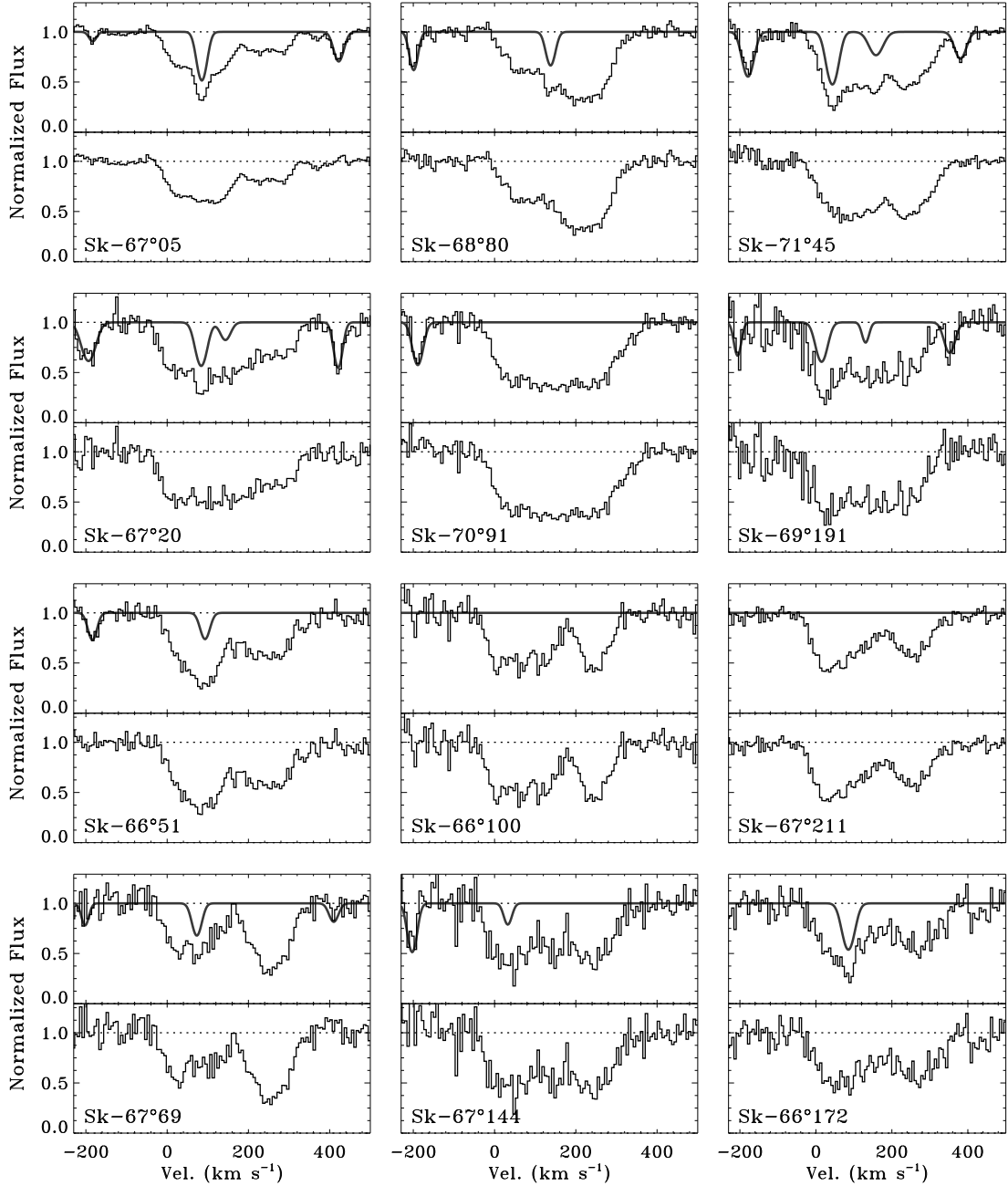}
\caption{Continuum normalized interstellar \ion{O}{6} absorption profiles 
for the sight lines being considered in this work.  For each star two
panels are shown: the normalized \ovi\ profiles with a model for the
contaminating H$_2$ absorption overplotted ({\em top}) and the
normalized profiles after division by the H$_2$ absorption model ({\em
bottom}).  These profiles are plotted after rebinning to 20 m\AA, or
$\sim6$ km s$^{-1}$, per pixel.
\label{fig:profiles} }
\end{figure}

Figure \ref{fig:profiles} shows the continuum-normalized interstellar
absorption profiles of \ovi\ towards all of the target stars listed in
Table \ref{tab:log}.  The top panel shows the profiles with the model
\htwo\ absorption spectrum over-plotted; the bottom panel shows the
continuum-normalized profiles with the \htwo\ model divided out.  The
models include the Milky Way P(3) and LMC R(4) components, which were
estimated using the same approach as the contaminating LMC P(3) and
Milky Way R(4) absorption.  Though these outlying components do not
overlap the \ovi\ absorption in which we are interested, they
demonstrate the quality of our procedure for estimating the \htwo\
absorption strength.

For most of the sight lines studied in this work, the contamination of
the \ovi\ profile by molecular hydrogen absorption does not
significantly affect the LMC absorption component.  Notable exceptions
include the sight lines towards \sk68\dg80, \sk71\dg45, and to lesser
extents \sk67\dg20, and \sk69\dg191.  

The (6-0) R(0) transition of the HD molecule at 1031.912 \AA\ is
almost coincident with the rest wavelength of the strong \ovi\
transition.  For the sight lines towards the LMC stars studied here,
contamination of \ovi\ by this line is unimportant because the \htwo\
column densities, which are expected to be $\sim10^4$ to $10^5$ times
stronger, are low (see Tumlinson et al. 2001).  Furthermore, several
HD lines of similar strength to the 6-0 band transition exist in the
\fuse\ bandpass (see Sembach 1999) but are not seen in the spectra.  
We have, therefore, made no correction for HD contamination of the
\ovi\ absorption.

\subsection{Interstellar Absorption Line Measurements}
\label{subsec:measurements}

We measured equivalent widths and column densities of the interstellar
\ovi\ \wave{1031.926} following Sembach \& Savage (1992).  We estimated
interstellar column densities using the apparent optical depth method
(Savage \& Sembach 1991).  The apparent optical depth, $\tau_a(v)$, is
an instrumentally-blurred version of the true optical depth of an
absorption line, given by
\begin{equation}
\tau_a(v) = - \ln \left[ I(v)/I_c (v) \right]
	\label{eqn:tauv}
\end{equation}
where $I_c(v)$ is the estimated continuum intensity and $I(v)$ is the
observed intensity of the line as a function of velocity.  This is
related to the apparent column density per unit velocity, $N_a(v)$ [${
\rm atoms \ cm^{-2} \ (km \ s^{-1})^{-1}}$], by
\begin{equation}
N_a(v) = \frac{m_e c}{\pi e^2} \frac{\tau_a (v)}{f \lambda} = 3.768
\times 10^{14} \frac{\tau_a (v)}{f \lambda(\mbox{\AA})},
\end{equation}
where $\lambda$ is the wavelength in \AA, and $f$ is the atomic
oscillator strength.  In the absence of {\em unresolved} saturated
structure the $N_a(v)$ profile of a line is a valid,
instrumentally-blurred representation of the true column density
distribution as a function of velocity, $N(v)$.  In the case of
interstellar \ovi\ absorption, the absorption profiles are typically
broad enough to be fully-resolved by the \fuse\ spectrographs (e.g.,
see Savage et al. 2000).  In the few cases where the weaker \ovi\
1037.617 \AA\ line is clean enough to allow a comparison with the
stronger line, no evidence for unresolved saturation has been found.

Table \ref{tab:columns} lists the derived \ovi\ the equivalent widths
and column densities for the LMC material.  The error estimates in
Table \ref{tab:columns} include contributions from statistical
(photon) uncertainties and uncertainties in the continuum polynomial
coefficients (see Sembach \& Savage 1992).  The effects of a $2\%$
flux zero-level uncertainty are $\sim0.01$ dex.

Also given in Table \ref{tab:columns} is the range of velocities over
which the absorption profile was integrated to yield the measurements
given for the LMC material.  The separation between Milky Way and LMC
absorption is not always clearly demarcated in the spectra (see Figure
\ref{fig:profiles}).  The separation is quite clear for the sight lines 
towards Sk--67\dg05, Sk--67\dg69, Sk--66\dg100, \sk67\dg211, and
\sk66\dg172.  For the remaining sight lines, the derived properties of
the LMC \ovi\ absorption are somewhat dependent on the limits of the
velocity integration.  For the cases where the separation between the
Milky Way and LMC absorption is not clear, we have adopted a standard
lower velocity of $v = +175$ \kms, which is roughly typical of the
sight lines for which the separation is relatively clear.  For all of
our measurements of the LMC \ovi\ column densities, we have also
derived these column densities at velocities $\pm20$ \kms\ from the
assumed lower velocity limit.  The difference between the columns
derived with these limits and that derived using the limits quoted in
Table \ref{tab:columns} is then treated as a $1\sigma$ uncertainty and
added in quadrature to the statistical and continuum placement
uncertainties discussed above.  The estimated uncertainties derived
using these $\pm20$ \kms\ integration limits were in the range 0.02 to
0.09 dex ($\approx 5\%$ to $23\%$), with an average of $\sim0.05$ dex
($\sim12\%$).

\section{Content and Distribution of \ion{O}{6} within the LMC}
\label{sec:columns}

In this section we discuss the total column densities and distribution
of interstellar \ovi\ within the LMC.  The stars in this work were
chosen to have early spectral types and well-behaved \ovi\ P Cygni
profiles.  Many of the stars in the sample were observed in order to
study their stellar properties.  As such we have not selected these
sight lines based solely upon the properties of the LMC ISM in these
directions.  Even so, these sight lines probe a variety of physical
regions within the LMC.  We discuss the physical environments of the
individual sight lines and compare the general distribution of \ovi\
absorption with that of the gas traced by
\halpha\ and X-ray emission.

\subsection{Total Line-of-Sight \ion{O}{6} Column Densities}

\begin{figure}
\epsscale{1.0}
\plotone{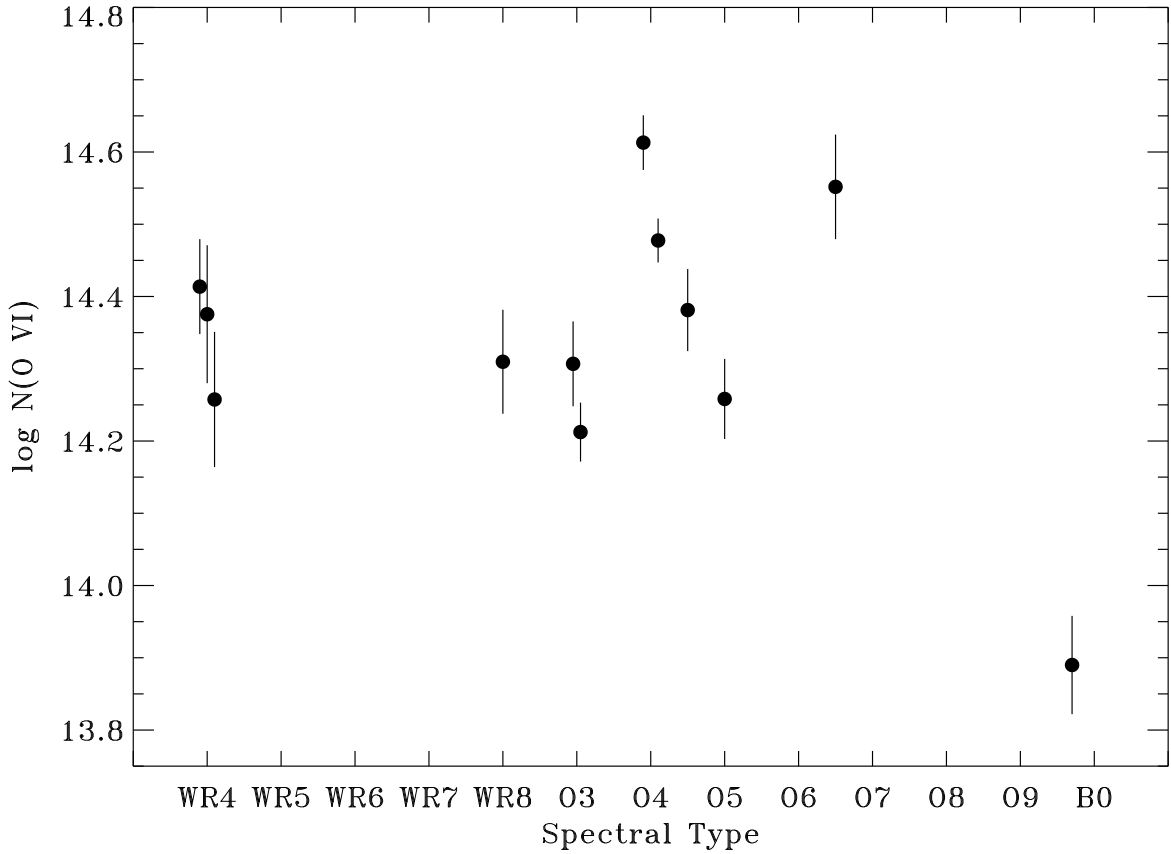}
\caption{Interstellar \ovi\ column in the LMC as a function of spectral 
type of the background probe.  We see no apparent trends in the \ovi\
column density with spectral type of the probe star, suggesting the
local effects of the stars themselves may play a minor role if any in
providing the observed \ovi\ columns.
\label{fig:sptype}}
\end{figure}

The integrated LMC \ovi\ column densities span the range $\log
N(\mbox{\ovi}) \sim13.9$ to 14.6 (in units of atoms \column), i.e., a
factor of $\sim5$.  We have detected \ovi\ absorption along all of the
sight lines presented in this work.  This is suggestive of a high
surface covering factor of \ovi\ in the LMC.  We are hesitant to place
too much emphasis on this result given that we have selected only
early O and WR stars as background probes.  The effects of the high
wind luminosities of these stars on their local environments could in
principle provide some of the observed \ovi\ column densities.  Figure
\ref{fig:sptype} shows the column density of LMC
\ovi\ as a function of the spectral type of the background star.
While there is no evidence for a dependence on spectral type, we
sample a limited range of spectral types.

The statistics of the \ovi\ column densities along the sight lines
presented in this work are summarized in Table \ref{tab:statistics},
including the straight and weighted means, standard deviation, and
median of the sample.  We give these statistics for the full sample of
12 stars shown in Table \ref{tab:columns} and for the sample without
the star \sk67\dg05.  The sight line to this star was discussed by
Friedman et al. (2000).  Because of its late spectral type (O9.7 Ib)
and relatively complex stellar continuum, we would not have included
it in our sample had it not been discussed in detail previously.  The
medians and unweighted means of the two samples are very similar.  We
will henceforth discuss the values from the entire sample, with the
caveat that one of our twelve sight lines may suffer from different
systematic effects than the others.

As shown in Table \ref{tab:statistics}, the average of the LMC column
density measurements is $\log \langle N(\mbox{\ovi}) \rangle = 14.37$
with a standard deviation of $\pm38\%$ ($^{+0.14}_{-0.21}$ dex).  The
logarithm of the median of the measurements is 14.38.  There is
significant variation in the column densities between the individual
sight lines.  The separations of the background probes used here range
from $0\fdg5$ to $5\fdg1$ ($\sim450$ to 4550 pc)\footnote{Throughout
this work we adopt the canonical distance of 50 kpc to the LMC.}.
Over the four smallest separations probed (450 to 510 pc), the
absolute differences in the logarithmic column densities are 0.15,
0.17, 0.20, and 0.37 dex, corresponding to linear ratios of 1.41,
1.48, 1.58, and 2.34.  The last value is for the
\sk67\dg05/\sk67\dg20 pair, which we point out because of the potential 
uncertainties associated with the former star.  We will discuss the
distribution of \ovi\ column densities more fully in \S
\ref{subsec:distribution}.

An obvious point of comparison for interpreting the column densities
of \ovi\ in the LMC is the Milky Way.  Savage et al. (2000) reported
the first \fuse\ observations of interstellar \ovi\ absorption in the
disk and halo of the Milky Way as seen in the spectra of 11 active
galactic nuclei (AGNs).  These authors report a patchy distribution of
\ovi\ within the Milky Way halo.  Defining the column density of 
\ovi\ projected onto the plane of the Galaxy by $ N_\perp(\mbox{\ovi})
\equiv N(\mbox{\ovi}) \sin |b|$, the sample of column densities in 
the Savage et al. (2000) sample has a mean $\log \langle N_\perp
(\mbox{\ovi}) \rangle = 14.29^{+0.14}_{-0.21}$ (std. dev.), with a
median of 14.21 and a full range $\log N_\perp (\mbox{\ovi}) = 13.80$
to 14.64 (a factor of $\sim7$ range).  In a plane-parallel
exponentially-stratified layer the quantity $N_\perp$ is simply the
product of the midplane density, $n_o$, and the exponential scale
height, $h$, of the distribution.  Adopting a midplane density of
$n_{\rm O \, VI} = 2 \times \e{-8}$ \percc\ based on \copernicus\
observations (Jenkins 1978), Savage et al. find implied exponential
scale heights in the range $h_{\rm O \, VI} \approx 1.0$ to 7.0 kpc.
Assuming an irregular distribution of absorbing clouds they derive a
scale height for the clouds of $\sim2.7\pm0.4$ kpc.

In order to compare the observed column densities in the LMC with
those observed looking out from our position in the Galactic disk
toward distant AGNs, it is useful to convert the LMC values to column
densities projected perpendicular to the plane of the LMC, $N_\perp$,
similar to the $N \sin |b|$ treatment of Savage et al. (2000).  In the
case of the LMC, that requires correcting the observed column
densities for the inclination of the LMC.  The inclination of the LMC
is typically defined in terms of the angle of the disk plane from the
plane of the sky, with values in the literature typically near $i
\approx 33^\circ$ (e.g., Westerlund 1997).  Hence, the correction
factor is not the sine, but rather the cosine of the inclination from
the plane of the sky.  We find $\log \langle N_\perp(\mbox{\ovi})
\rangle \equiv \log \langle N(\mbox{\ovi}) \rangle \cos i = 14.29^{+0.14}_{-0.21}$ with the log of the median value being 14.30.

The average column density of \ovi\ projected onto the plane of the
LMC in our sample of 12 sight lines is identical to that projected
onto the disk of the Milky Way, with the same degree of variation.  We
note that the same calculations excluding the sight line towards
\sk67\dg05 yield an average column density higher by $0.03$ dex
($7\%$) with a lower standard deviation. We will discuss the
implications of this comparison of the hot gas content of the Milky
Way and LMC more fully in \S \ref{sec:discussion}.

\subsection{Comparison with Other Views of the LMC}
\label{subsec:distribution}

The ISM of the LMC has been studied extensively over a range of
wavebands.  Given its proximity, these studies have provided a wealth
of information on the warm ($\sim10^4$ K) ionized gas traced through
its \halpha\ emission and hot ($\ga 10^6$ K) ionized material traced
through its X-ray emission.  Surveys of the former, in particular,
have been used to identify large-scale structures in the ISM of the
LMC that trace the feedback of energy from massive stars into the ISM
(Henize 1956; Davies et al. 1976; etc.).  \halpha\ imaging
observations have revealed the presence of ionized structures on
almost all scales, including \HII\ regions, supernova remnants (SNRs),
superbubbles, and ``super-giant shells'' (SGSs).  The latter two
classes of \halpha -emitting structures are the most prominent with
projected diameters of $\sim100$ to $\sim1000$ pc.

\subsubsection{Comparison with \halpha\ Distribution}

Many of the stars in our \fuse\ sample are projected onto prominent
\halpha -emitting structures.  Table \ref{tab:environs} lists the
identification of any nebulosity and describes the general
interstellar environments for each sight line.  The identifications
and descriptions rely heavily on previous studies.  Examining Table
\ref{tab:environs} one can see a range of the general interstellar
environments probed, including sight lines that are projected onto
very faint \halpha\ emission associated with diffuse \HII\ regions as
well as objects that are projected onto prominent shells that have
been identified as superbubbles.  Several sight lines also probe the
SGS LMC 4.  Thus, while our sample of sight lines is small, it seems
to probe a wide variety of physical regions.

\begin{figure}
\epsscale{1.0}
\plotone{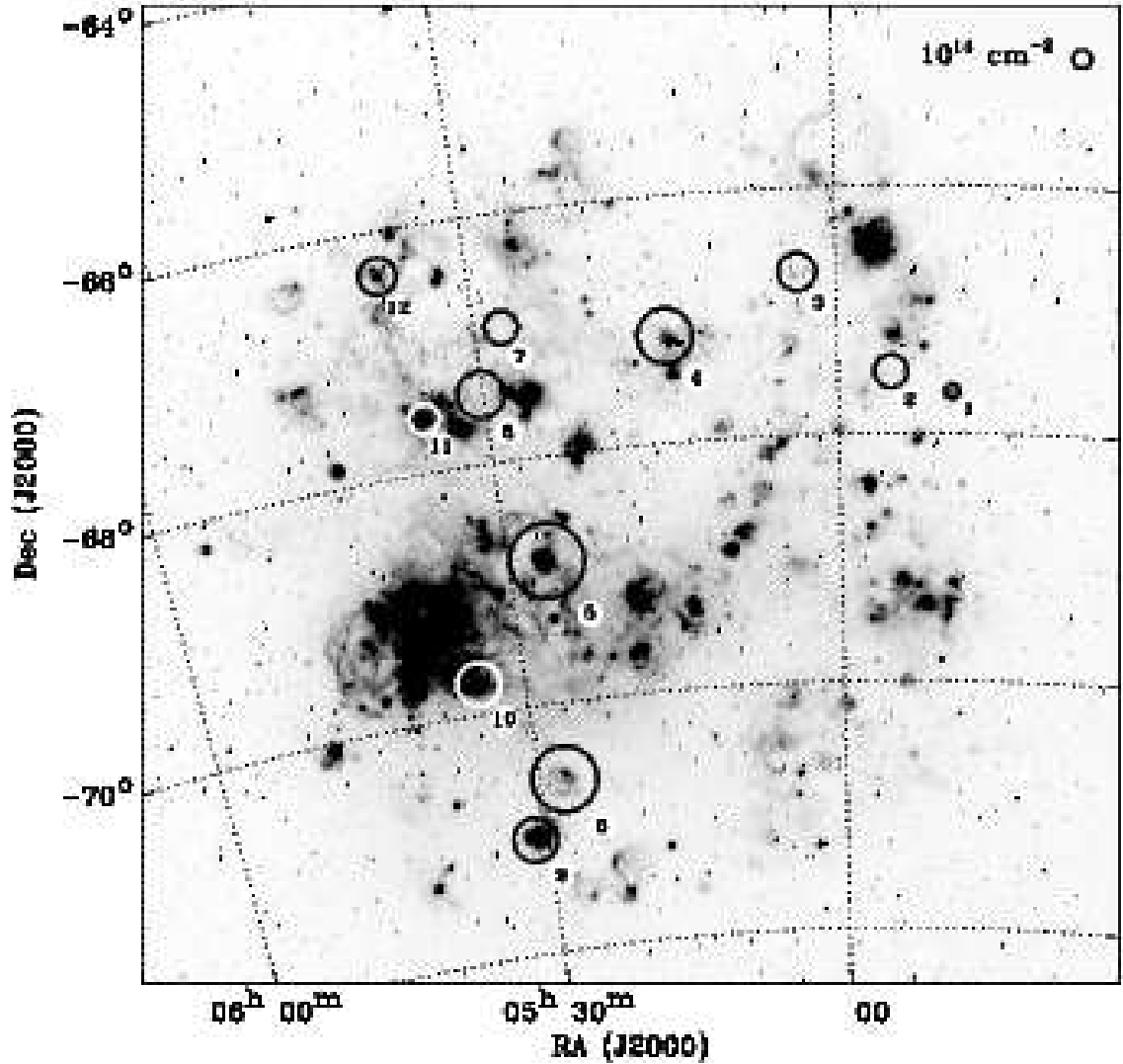}
\caption{H$\alpha$ image of the LMC (Gaustad et al. 2001) with the 
positions of the probe stars marked.  Darker regions correspond to
brighter \protect\halpha\ emission.  The radius of the circle marking
each probe star is linearly proportional to the column density of
interstellar \protect\ovi\ at LMC velocities.  For scale, a circle
corresponding to $N(\mbox{\protect\ovi})=10^{14}$ cm$^{-2}$ is given
in the upper right of the image.  The numbers given beside each circle
correspond to the identifications listed in Tables \ref{tab:columns}
and \ref{tab:environs}.  Note that much of the apparent emission near
Sk--67$^\circ \, $69 ($\#4$) is caused by an imperfectly-subtracted
foreground star.
\label{fig:halpha}}
\end{figure}

Figure \ref{fig:halpha} presents a representation of the interstellar
\ovi\ column densities as a function of position on the sky overlaid 
on an \halpha\ image of the LMC (Gaustad et al. 2001).  The directions
probed by our \ovi\ measurements are marked with circles whose radii
are linearly proportional to the integrated LMC \ovi\ column densities
along these lines of sight.  Hence, Figure \ref{fig:halpha} shows a
``map'' of the \ovi\ column density across the face of the LMC.  A
number is given next to each circle that denotes each sight line with
the ID given in Table \ref{tab:columns}, and the circle in the upper
right of the figure gives the scale for $N(\mbox{\ovi})=10^{14}$
\column. More detailed views of the \halpha -emitting gas projected
near these sight lines can be found in Danforth et al. (2002).

Figure \ref{fig:halpha} gives an interesting view of the \ovi\
distribution within the LMC.  This figure demonstrates that relatively
strong \ovi\ absorption is present in regions where the ISM traced by
\halpha\ is relatively quiescent.  For example, the lines of sight to
\sk67\dg20, \sk66\dg51, and \sk67\dg69 (numbers 2 through 4 in 
Figure \ref{fig:halpha}) all probe regions of only relatively diffuse
\halpha\ emission in quiescent regions of the galaxy (the bright spot
near the last of these is due to an incompletely-subtracted foreground
star), though their \ovi\ column densities exceed $10^{14}$ \column.

It can also be seen from Figure \ref{fig:halpha} that several stars
(e.g., \sk66\dg100, \sk67\dg144, and \sk66\dg172 -- numbers 7, 8, and
12) probe the periphery or interior of the SGS LMC 4, a 1400 pc
diameter ring delineated by \HII\ regions and encompassing an \HI\
void (Kim et al. 1998) centered at $\alpha_{\rm J2000} \approx 05^h\,
31^m\, 33^s$; $\delta_{\rm J2000} \approx -66^\circ\, 40\arcmin\,
28\arcsec$.  Furthermore, those lines of sight south of $\delta_{\rm
J2000} = -68^\circ$ are all projected onto previously-identified
``supershells'' in \halpha\ and X-ray studies (e.g., Dunne, Points, \&
Chu 2001).  While two of these four sight lines exhibit the largest
column densities in our sample (\sk68\dg80 and \sk71\dg45 -- numbers 5
and 9, which probe N144 and N206, respectively), the other two have
column densities consistent with the median of the sample.  

\begin{figure}
\epsscale{1.0}
\plotone{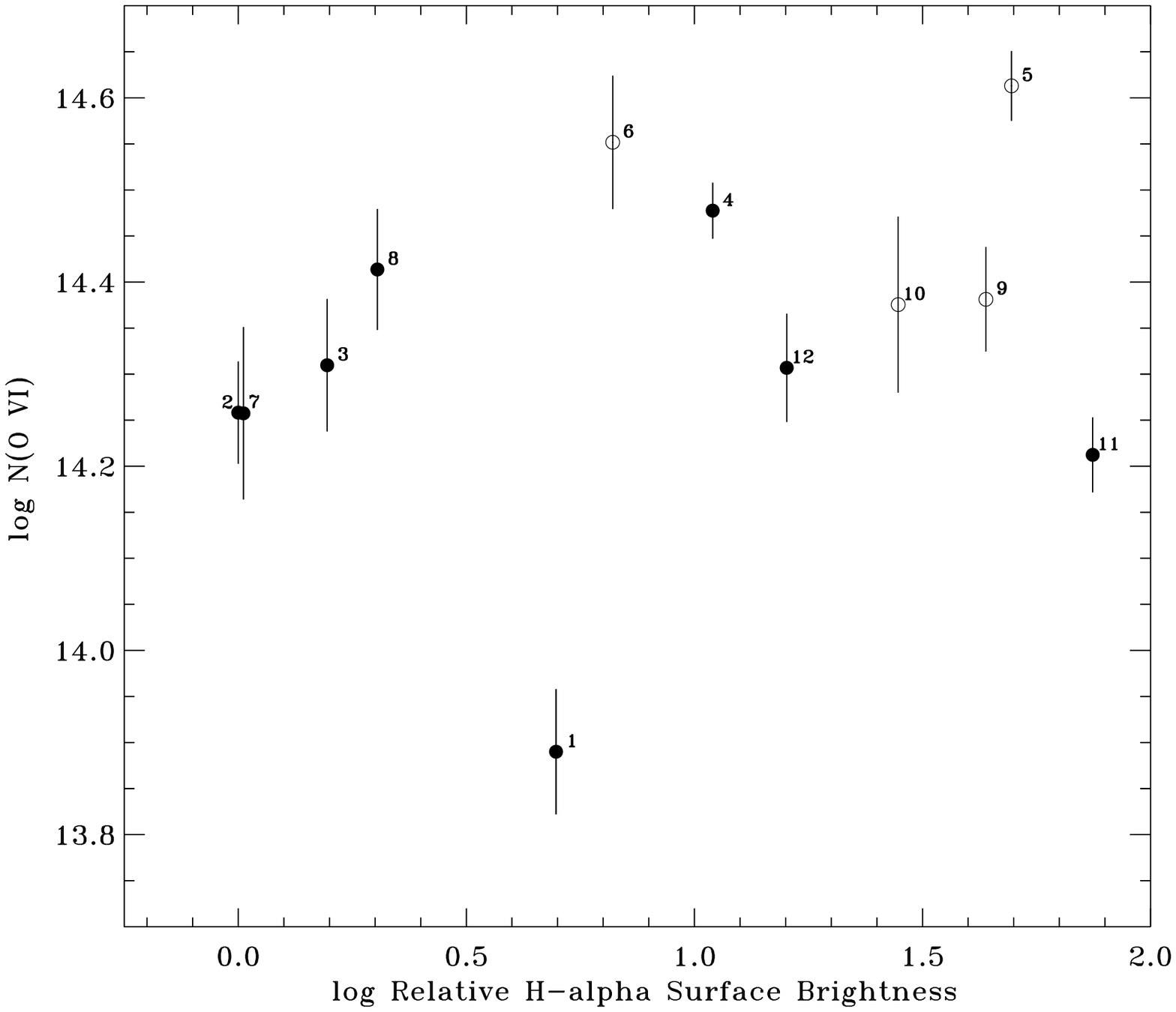}
\caption{Interstellar \protect\ovi\ column density versus 
relative \protect\halpha\ surface brightness (averaged over
$6\farcm6\times6\farcm6$ -- $100\times100$ pc$^2$ -- boxes) for the
sight lines studied in this work.  The numbers given beside each point
correspond to the identifications listed in Tables
\protect\ref{tab:columns} and \protect\ref{tab:environs}.  Open 
circles denote directions towards identified superbubbles (see Table
\protect\ref{tab:environs}).  
\label{fig:hao6}}
\end{figure}

The \ovi\ column densities observed along our 12 sight lines seem to
have little relation to the observed morphological structure of
\halpha\ emission (Gaustad et al. 2002) in these directions.  Furthermore, 
there is no relationship between the observed \ovi\ column densities
and the local star formation rate as traced by the intensity of the
\halpha\ emission.  Figure \ref{fig:hao6} shows the observed \ovi\
column densities as a function of the relative \halpha\ intensity
integrated over $6\farcm6\times6\farcm6$ (100x100 pc$^2$) regions
surrounding the target stars.  The range in \halpha\ surface
brightness measurements is much larger than the range in \ovi\ column
densities.

The sight lines probing superbubbles and supergiant shells have \ovi\
column densities quite similar to those projected onto more quiescent
regions of the galaxy, and our derived \ovi\ column densities have no
observable relationship to the local \halpha\ intensity within the
LMC.  Thus, there is no evidence for large enhancements in the
\ovi\ column densities associated with large-scale structures or high
levels of local star formation.  As discussed in \S
\ref{sec:discussion}, this is consistent with a scenario in which much
of the observed \ovi\ resides in an extended, but patchy, distribution
such as a galactic ``corona.''

\subsubsection{Comparison with X-ray Distribution}

Hot ionized material in galaxies can also be revealed through
observations of its X-ray emission.  The general distribution of X-ray
emission in the LMC has been studied with several instruments
including the {\em Einstein Observatory} (Wang et al. 1991) and {\em
ROSAT} (Snowden \& Petre 1994).  These studies have discussed the
observations in the context of the diffuse, hot ISM in the LMC.  Many
studies detailing various aspects of individual regions of the LMC
have also been published.  The most relevant for comparison with this
work are recent summaries of X-ray emission from superbubbles (Dunne
et al. 2001) and large-scale diffuse regions, including SGSs (Points
et al. 2001).  These works include discussions of diffuse X-ray
emission from some of the structures (regions) probed by our
absorption line measurements.

\begin{figure}
\epsscale{1.0}
\plotone{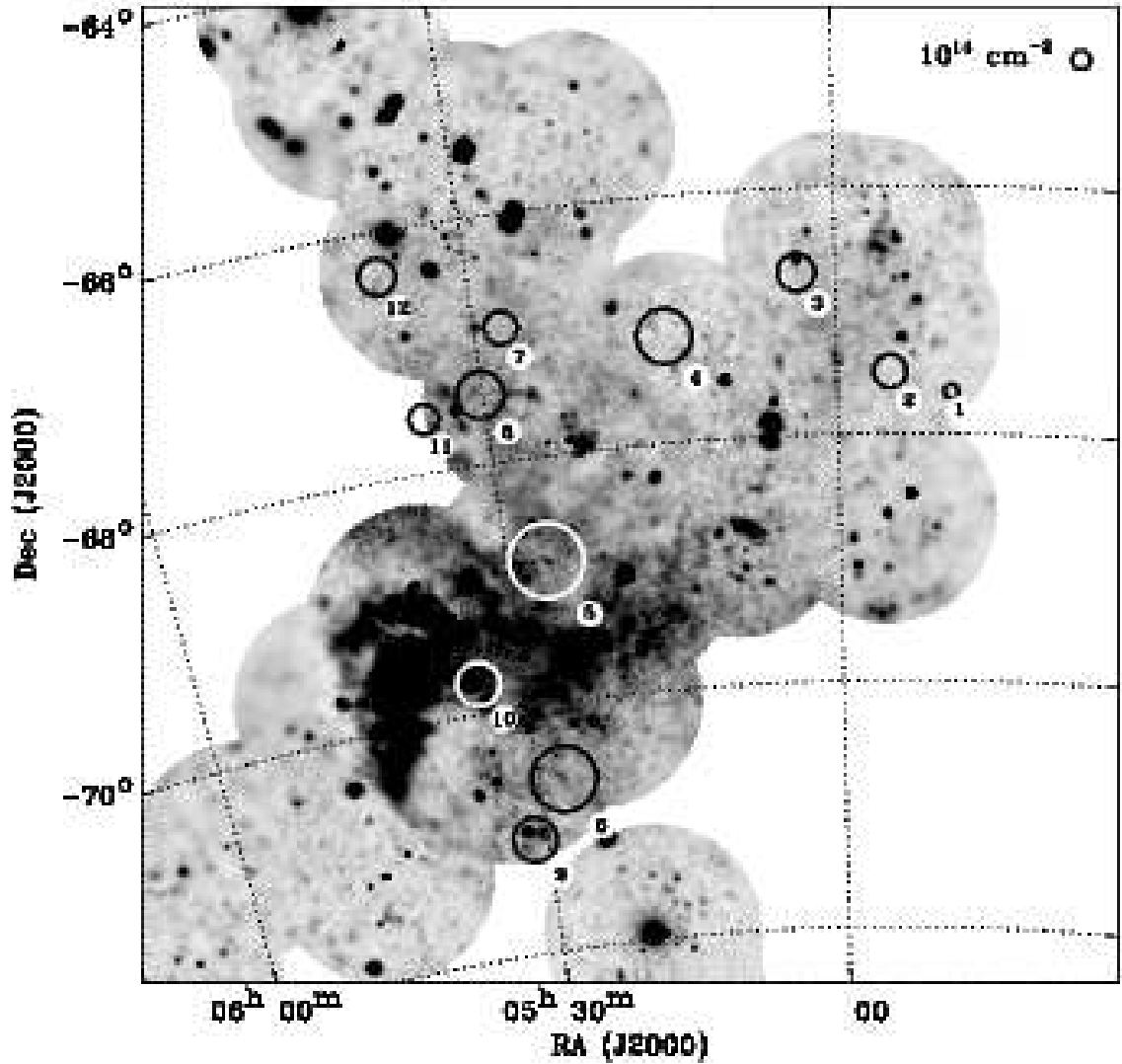}
\caption{\protect\rosat\ PSPC mosaic of the LMC (Snowden \& Petre 1994) 
with the positions of our probe stars marked.  This image shows the
R4--R7 band image, which covers the 0.5--2.0 keV energy range.  The
angular resolution varies with position in this adaptively-smoothed
image, but is always larger than the 20\arcsec\ limit for the PSPC.
Darker regions correspond to brighter X-ray emission.  As in Figure
\protect\ref{fig:halpha}, the radius of the circle marking the
position of each probe star is linearly proportional to the column
density of interstellar \protect\ovi\ at LMC velocities, with a scale
given in the upper right.  The numbers given beside each circle
correspond to the identifications in Tables \protect\ref{tab:columns}
and \protect\ref{tab:environs}.
\label{fig:rosat}}
\end{figure}

In general it is difficult to study the soft X-ray emission from the
LMC given the relatively large foreground photoelectric opacity.
Unfortunately, this is the waveband that is most likely to correspond
to the material being traced by our \ovi\ absorption line
measurements.  However, the harder X-ray emission traces gas at
relatively high temperatures ($10^6 - 10^7$ K) which will presumably
cool through the temperatures ($\la 3\times 10^5$ K) at which \ovi\ is
expected to be most abundant.  Therefore, one might expect at least a
loose connection between these two tracers of hot ionized matter.
Figure \ref{fig:rosat} shows the \rosat\ PSPC mosaic of the LMC from
Snowden \& Petre (1994) with the \ovi\ columns from this work
displayed as in Figure \ref{fig:halpha}.  The X-ray mosaic makes use
of the R4 to R7 filters of the PSPC, which are sensitive to $\sim0.5$
to 2.0 keV photons, and has been adaptively smoothed, as discussed in
Snowden \& Petre.  This energy range is non-uniformly affected by
photoelectric absorption by LMC and Milky Way gas.  Point sources
(e.g., X-ray binaries and background AGNs) have not been removed.

The main purpose of presenting this mosaic in comparison with our
\ovi\ measurements is to demonstrate that the sight lines probed in
this work do not, in fact, probe regions of the LMC that are unusually
X-ray bright.  Our sample includes observations of a star projected
onto the superbubble N144 and three sight lines projected onto the SGS
LMC 4.  Recent work on archival \rosat\ PSPC data have shown that
these regions do not have excessively high X-ray surface brightnesses
(Dunne et al. 2001; Points et al. 2001).  Indeed, supernova remnants
are the largest contributors to the interstellar X-ray emission of the
LMC, while superbubbles and SGSs are collectively more than a factor
of two lower (Points et al. 2001).  In general our sight lines do not
probe regions where the amount of hot gas is atypically large for the
LMC.

\begin{figure}
\epsscale{1.0}
\plotone{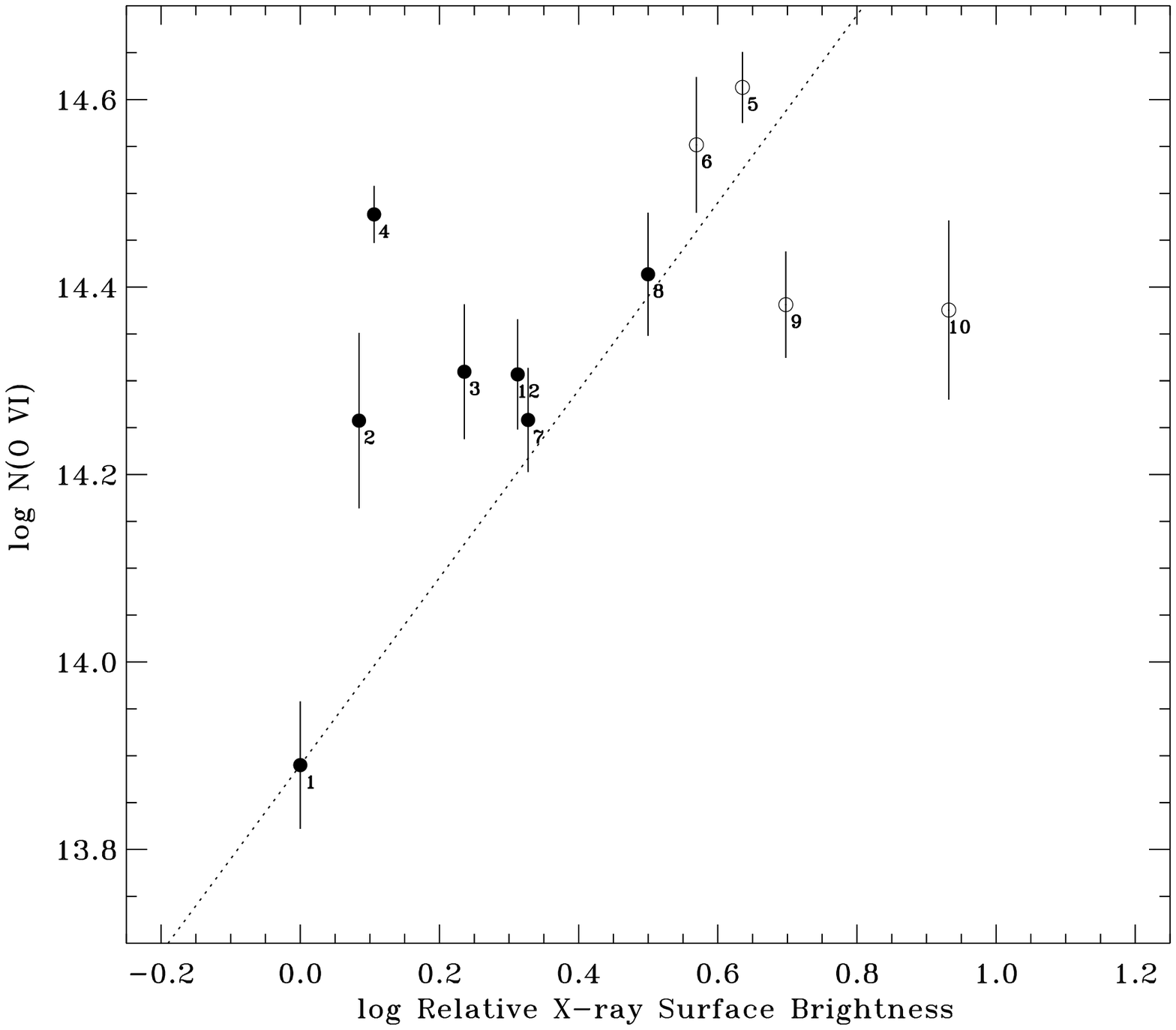}
\caption{Interstellar \protect\ovi\ column density versus 
relative \protect\rosat\ 0.5 to 2.0 keV PSPC surface brightness
(averaged over $6\farcm6\times6\farcm6$ -- $100\times100$ pc$^2$ --
boxes) for the sight lines studied in this work.  The numbers given
beside each point correspond to the identifications listed in Tables
\protect\ref{tab:columns} and \protect\ref{tab:environs}.  Open circles 
denote directions towards identified superbubbles (see Table
\protect\ref{tab:environs}).  The dashed line corresponds to a linear
one-to-one relationship between the \protect\ovi\ column density and
X-ray surface brightness.  Note that the sight line towards
Sk--67$^\circ \, $211 is not covered by the \protect\rosat\ mosaic
and, therefore, is not represented in this figure.  The lowest
\protect\ovi\ column density is seen towards Sk--$67^\circ\,$05
($\#1$).  The continuum placement for this star is less certain than
for the other objects (see Friedman et al. 2000).
\label{fig:xo6}}
\end{figure}

Figure \ref{fig:xo6} shows the observed \ovi\ column densities
compared with the relative X-ray surface brightness from \rosat\
(Snowden \& Petre 1994) for each sight line.  The X-ray surface
brightnesses were calculated for boxes $6\farcm6$ (100 pc) on a side,
centered on the star of interest.  No correction for overlying
photoelectric opacity has been made in deriving the X-ray
brightnesses.  A linear relationship of slope unity between the two
datasets is shown with the dashed line in the plot.  We note that if
the lowest point is left out (representing the sight line towards
\sk67\dg05), this diagram becomes a scatter plot with no discernible
relationship between the \ovi\ column density and X-ray surface
brightness measurements.  Indeed, a Spearman's rank-order correlation
test (Press et al. 1992) cannot rule out the null hypothesis that the
\ovi\ columns and X-ray surface brightnesses are uncorrelated at the
$2\sigma$ level.  The significance of any correlation between the
quantities is small.  Given the very different way in which these
measurements are made, the differing systematics to which they are
subject, and the different gas traced by each approach, it would be
surprising to find a significant correlation.

The four sight lines in our sample probing superbubbles are denoted
with open circles in Figure \ref{fig:xo6}, while the other sight lines
are marked with filled circles.  The superbubble sight lines have the
highest four X-ray surface brightnesses, though their \ovi\ column
densities are not so easily separated from the other measurements.
This may suggest that local effects are more important for determining
the hard X-ray brightness than they are for determining the \ovi\
column densities.  However, the \rosat\ ``beam'' is much larger than
that probed by our absorption line measurements, and we do not know
that the stars used as background probes do indeed lie behind the gas
associated with the superbubbles.

\section{Gross Kinematic Properties of \ion{O}{6} within the LMC}
\label{sec:kinematics}

Figure \ref{fig:profiles} demonstrates two items of note regarding the
kinematics of the observed \ovi\ absorption: 1) There is a large
degree of variation in the observed kinematic profiles of the Milky
Way/LMC absorption complex; and 2) Separating the Milky Way and LMC
components is non-trivial in some cases.  The identification of \ovi\
absorption in the velocity range $150 \la v \la 200$ \kms\ with either
of the two galaxies is particularly difficult.  At lower velocities,
profiles such as those seen towards \sk66\dg100, \sk67\dg69, and
\sk67\dg211 strongly suggest a Milky Way (including high- and
intermediate-velocity cloud) origin.  However, there is not
easily-identifiable separation between Milky Way and LMC absorption in
profiles such as those towards \sk70\dg91 and
\sk67\dg20.   This ambiguity between the absorption from the two galaxies   
makes it difficult to definitively discuss the kinematics of the
absorption from the LMC.  In particular, some of the material in this
intermediate range could be caused by outflows from the LMC with
velocities of $\sim100-150$ \kms.  If this is the case, it is an
important aspect of the LMC \ovi\ kinematics, although one that cannot
be addressed by the current observations.

Even in the presence of such ambiguities, there is useful information
on the kinematics of the LMC \ovi\ within the observed absorption
profiles.  We discuss two particular aspects of the LMC \ovi\
kinematics in this section: a comparison of the \ovi\ velocity
profiles with those of the lower-ionization material traced by \feii;
and the kinematic properties of the \ovi\ in directions where there is
a clear separation between the Milky Way and LMC absorption.

\begin{figure}
\epsscale{0.75}
\plotone{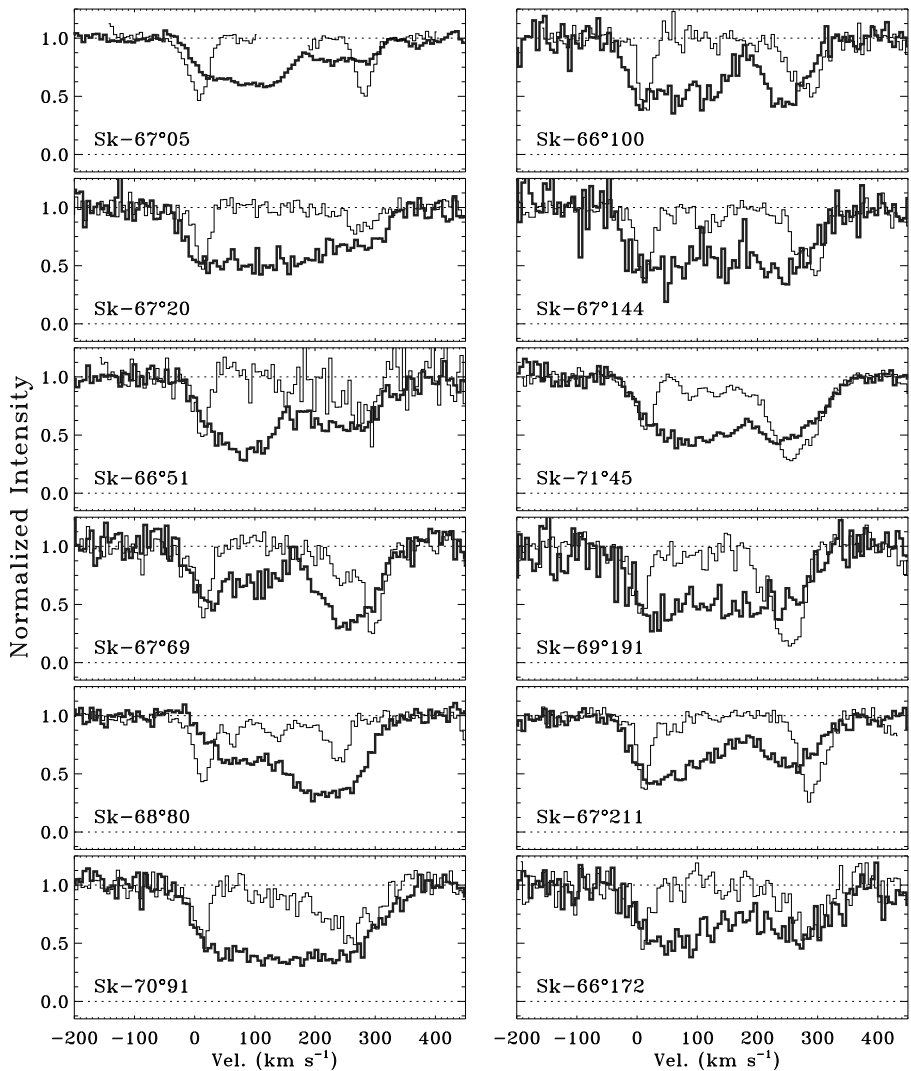}
\caption{The observed absorption line profiles of \protect\ion{O}{6} 
$\lambda 1031.926$ (thick line) and \protect\ion{Fe}{2} $\lambda
1125.448$ (thin line) towards the LMC stars studied in this work.  The
\protect\ion{Fe}{2} absorption components associated with the Milky 
Way and LMC are seen to be significantly narrower than the
corresponding \protect\ion{O}{6} profiles.  Also present to some
degree along all sight lines is gas associated with intermediate- and
high-velocity clouds at $v\sim+65$ and +125 km s$^{-1}$, respectively.
The effective offsets of LMC \protect\ion{O}{6} and
\protect\ion{Fe}{2} absorption may be caused by rotational effects or 
by material outflowing from the disk.  The \protect\ion{Fe}{2}
profiles towards Sk--67\dg05 for velocities intermediate between the
Milky Way and LMC absorption are not shown because of the presence of
confusing stellar absorption at these velocities.
\label{fig:phases}}
\end{figure}

Figure \ref{fig:phases} shows a comparison of the \ovi\
\wave{1031.926} absorption line profiles (cleaned of \htwo) with the 
\feii\ \wave{1125.448} profiles.  The latter trace low-ionization gas 
associated with the relatively high column density warm neutral and
warm ionized media in the Milky Way and LMC.  Presumably most of the
material in these cooler phases ($\sim10^3-10^4$ K gas) resides in or
near the disk of each galaxy.  Examining the \feii\ profiles, one can
see clear signatures of Milky Way, intermediate- and high-velocity
clouds (IVCs and HVCs), and LMC absorption.  All of the sight lines in
our sample show evidence for either IVC and/or HVC absorption.  Though
these components are not always readily visible in the \feii\
\wave{1125.448} profiles, the stronger \feii\
1144.938 \AA\ and \ion{O}{1} 1039.230 \AA\ transitions show this to be
the case (see the profiles presented in Danforth et al. 2002).  The
LMC often exhibits multiple components or complex absorbing structure.
The stronger \feii\ 1144.938 \AA\ transition often shows much lower
column density components that are not visible in the
\feii\ \wave{1125.448} profiles.

The individual \feii\ components are more easily separated than those
in the \ovi\ profiles because the widths of typical \feii\ absorption
complexes is significantly lower than those seen in \ovi\ absorption.
Good examples of this effect are seen in the spectra of
\sk67\dg69, \sk68\dg80, and \sk67\dg211.  The differences between 
the \feii\ and \ovi\ absorption could be caused by larger thermal
broadening within the hot ISM traced by the \ovi\ absorption.  While
this is in part the case, we show below that the individual \ovi\
absorbing complexes are much broader than expected from pure thermal
broadening.

The \ovi\ profiles almost always extend to lower velocities than the
\feii\ absorption.  Furthermore, in cases where the LMC \ovi\
absorption is well separated from the Milky Way, it tends to be
somewhat symmetric and centered at velocities well below those of the
\feii\ absorption complexes.  The \feii\ absorption is often aligned 
with the high-velocity edge of the \ovi\ absorption.  Thus the \ovi\
often appears to be kinematically decoupled from the \feii, which
traces material predominantly associated with the disk of the LMC.

A similar effect, but opposite in sign, can be seen when comparing the
\ovi\ and \feii\ absorption from the Milky Way.  The \feii\ typically
traces the low-velocity edge of the \ovi\ absorption, the latter
extending to much higher velocities.  This may in part be caused by
the stronger influence of the IVCs and HVCs on the \ovi\ profile.
However, because Galactic rotation Doppler shifts distant gas to
positive velocities in this direction, much of this effect could also
be caused by the larger vertical extent of \ovi\ compared with \feii.

Several sight lines show enough separation between the Milky Way and
LMC \ovi\ absorption that the detailed kinematic properties of the LMC
absorption can be ascertained.  We have fit Gaussian components to the
apparent optical depth profiles for several lines of sight, deriving
the average velocity, breadth, and peak optical depth for the LMC
absorption seen towards these stars.  We assume that the LMC
absorption can be approximated by a single Gaussian component.  In
most cases this gives a reasonable approximation to the observed
absorption profiles.  In these cases we have also fit the Milky Way
absorption to account for any minor blending with the LMC absorption.
We do not report the Milky Way results here, particularly because
there is a large degree of ambiguity when fitting the \ovi\ profiles
with multiple Gaussians.

The absorption profiles fit in this way include those towards the
stars \sk67\dg69, \sk66\dg100, \sk67\dg211, and \sk66\dg172.  Further
decomposition of the profile towards \sk71\dg45 is discussed by
Lauroesch et al. (2002).  The most important aspect of these fits is
the derived dispersion of the LMC \ovi\ absorption.  For the sight
lines listed we derive $\sigma = 37$, 30, 41, and 48 \kms,
respectively.  The corresponding FWHM ($\Gamma$) values are $\Gamma =
87$, 71, 96, and 113 \kms.  This range of widths implies temperatures
for the absorbing regions $T \leq 2\times10^6$ to $\leq 5\times10^6$
K.  If the observed gas is closer to the temperature at which the
abundance of O$^{+5}$ peaks in collisional ionization equilibrium
($T\sim3\times10^5$ K), most of the observed breadths of these
profiles must be caused by non-thermal broadening mechanisms,
including turbulence, multiple velocity components, and large-scale
dynamical effects.  Given the smoothness of the profiles observed at
high signal-to-noise ratios, it seems likely that if this breadth is
due to multiple absorbing components along the line of sight then
several such components may be required.

Savage et al. (2002) have measured the apparent breadths of the
absorption from the \ovi\ within the halo (or thick disk) of the Milky
Way as seen towards a large number ($\sim85$) of extragalactic
sources.  These authors find $\sigma \approx 16$ to 65 \kms\ ($\Gamma
\approx 37$ to 153 \kms) with a median of $\sigma \approx 39$ \kms.  
The observed breadths towards lower latitude sources may be affected
by Galactic rotation, although many of their sight lines are at high
enough Galactic latitudes that such effects are unimportant.  The
median breadth from these large number of measurements through the
Galactic halo is essentially equivalent to that of the four LMC
measurements (median $\sigma = 39$ \kms).  For comparison, Jenkins
(1978) has discussed the breadths of nearby ($d\la1000$ pc) disk
\ovi\ absorption as observed by \copernicus; he found an average
dispersion of $\sigma \approx 29$ \kms.

We have also fit Gaussian profiles to the observed \feii\ absorption
for the four sight lines listed above to determine the central
velocities of the LMC \feii.  Of interest in these cases is the
relative difference between the average velocities of the \ovi\ and
\feii\ absorption.  For this small sample of sight lines, we find that
an average velocity difference, $\Delta v \equiv v_{\rm O \, VI} -
v_{\rm Fe \, II} = -32$ \kms.  The central velocity of the
\ovi\ is on average shifted by $-32$ \kms\ relative to central 
velocity of the \feii\ along these lines of sight.  Because the \ovi\
is significantly broader than the \feii, much of the \ovi\ absorption
occurs at velocities significantly lower than the bulk of the \feii\
absorption (which can be seen in Figure \ref{fig:phases}).  This
average velocity separation may not be typical of all of the sight
lines since we have examined only those directions that show a clear
separation between the Milky Way and LMC \ovi\ absorption.

We note that our observations cannot be used to test the existence of
high-velocity outflows from the disk of the LMC.  Material separated
from the LMC disk velocity by $v - v_{\rm disk} \la -75$ \kms\ is
confused with Galactic ISM, IVC, and HVC absorption.  In almost all of
the sight lines there is \ovi\ absorption at velocities shortward of
the \feii\ absorption associated with the LMC disk (but longward of
the $\sim+150$ to $+175$ \kms\ separation between LMC and HVC
material).  \ovi\ absorption is usually not detected at positive
velocities relative to the \feii\ from the LMC disk.  The two
exceptions to this are \sk68\dg80 and perhaps \sk70\dg91.  The \feii\
profile towards the latter object shows several LMC components with
similar strength, making the identification of the ``disk'' component
somewhat ambiguous.  Towards \sk68\dg80 \ovi\ absorption extends to
velocities $\sim+60$ \kms\ higher than the \feii\ 1125.448 \AA\
absorption.  A very weak component can be seen in the \feii\ 1144.938
\AA\ line at $v\sim+275$ \kms, though the bulk of the \feii\ absorption 
in this direction is centered at $v\sim+240$ \kms.  The extended high
positive-velocity absorption seen in the \ovi\ profile towards
\sk68\dg80 may be evidence for infalling material along this sight line.
However, at least 10 of the 12 sight lines studied in this work reveal
no such absorption, suggesting infalling highly-ionized material is
rare in the LMC.

\section{Discussion: the Hot Corona of the LMC}
\label{sec:discussion}

Our observations of interstellar \ovi\ associated with the LMC reveal
that this tracer of hot, collisionally-ionized gas is present in
relatively large quantities across the whole face of the LMC, though
with a large degree of patchiness.  Lines of sight projected onto
superbubbles and supergiant shells have much the same column densities
as lines of sight projected onto quiescent regions (as indicated by
\halpha\ emission).  The velocity distribution of the LMC \ovi\
absorption is significantly different than that of lower-ionization
gas (e.g., \feii), being both much broader and present at lower
absolute velocities.

The average column density projected onto the plane of the LMC
($N_\perp$) is equivalent to that projected onto the plane of the
Milky Way, as determined by measurements of the Galactic halo towards
extragalactic sources (Savage et al. 2000).  The dispersion in the
measurements of $N_\perp$ in the LMC and Milky Way is also the same.
The kinematics of the highly-ionized gas traced by
\ovi\ in the LMC seems to be very similar to that seen through the halo 
of the Milky Way, although our ability to accurately measure the
kinematic profiles of the LMC absorption is limited.

There are several possible interpretations of these salient aspects of
the LMC \ovi\ absorption observed towards our sample of 12 stars.
However, given the striking similarities between the properties of the
LMC \ovi\ and the Galactic halo \ovi, the relatively small impact of
superbubbles on the total \ovi\ column densities, the existence of
significant quantities of \ovi\ along sight lines with little evidence
for local sources of highly-ionized material, and the general
decoupling of the \ovi\ kinematics from the disk material traced by
\feii, we favor an interpretation in which the LMC \ovi\ absorption
arises in a vertically-extended halo or corona similar to that seen
about the Milky Way (see Savage et al. 2000).  

\subsection{Alternative Interpretations}

Before exploring the \ovi\ corona interpretation in more detail, it is
worth mentioning alternatives to this scenario.  Sembach et al. (2000)
have presented observations of \ovi\ in the HVC system surrounding the
Galaxy.  Most (if not all) sight lines that pass through known \HI\
HVCs show \ovi\ absorption, including the Magellanic Stream (MS),
material tidally-expelled from the LMC/SMC system.  If the MS has a
component projected between the Sun and the LMC, the observed \ovi\ at
LMC velocities (or at the IVC and HVC velocities) could be associated
with this tidal structure.  For this to be true, however, the
ionization state of the Stream in these direction would be
significantly different than other directions probing this structure
(and than most other HVCs).  Given that the observed \feii\ absorption
can be associated with LMC disk material (see, e.g., Danforth \& Chu
2000) and is typically centered well away from the center of the \ovi\
absorption, relatively little low-ionization material could be
associated with such a putative Stream component.

Another possibility is that the \ovi\ absorption traces the
interaction between the gaseous coronae of the LMC and Milky Way.  The
presence of \ovi\ in the MS strongly suggests the Milky Way halo
extends to the distance of that structure, which is roughly
$\sim25-75$ kpc from the Sun (see the discussion by Sembach et
al. 2000).  It is therefore reasonable to expect that the Galactic
corona extends to the 50 kpc distance ($z\sim25$ kpc) of the LMC.
Interactions between the LMC corona (or other portions of this galaxy)
with the low-density, highly-ionized medium associated with the
extended distribution of hot gas of the Milky Way could, in principle,
provide for instabilities that cool the gas through the transition
temperatures at which O$^{+5}$ is abundant.  [de Boer et al. (1998)
have similarly suggested that the motion of the LMC through the
Galactic halo can trigger star formation through the compression of
LMC material along the leading edge of the LMC by the low density
Milky Way material.]  Given the large number of unknowns regarding the
structure of the ISM of the LMC and the extended Galactic halo, this
model remains largely unconstrained.  Unfortunately, at this point, it
is difficult to falsify either of the previous models.

\subsection{The Origins of \ovi\ in the Corona of the LMC}

The scenario we prefer is one in which the \ovi\ within the LMC is
distributed within the thin disk as well as in a more
vertically-extended thick disk or halo of the LMC.  Since our lines of
sight probe large pathlengths through the halo component, this
dominates over material associated with the disk, in a manner
analogous to observations of extragalactic sources from within the
disk of the Milky Way (Savage et al. 2000).  The presence of a
possible corona about the LMC has been discussed in the literature for
several years.  Early observations of \civ\ absorption towards LMC
stars with the \iue\ observatory revealed highly-ionized material at
velocities consistent with an LMC origin (Savage \& de Boer 1979,
1981; de Boer, Koornneef, \& Savage 1980; de Boer \& Savage 1980).
The early \iue\ measurements of LMC material were interpreted as
absorption from an extended hot corona.  That interpretation was
subsequently called into question (Chu et al. 1994; Feitzinger \&
Schmidt-Kaler 1982) because of the strong possibility that the high
ions observed were produced in the local environments of the observed
stars.  Wakker et al. (1998) used \hst\ observations of
\civ\ towards stars in quiescent regions of the LMC to argue that the
LMC did indeed contain a highly-ionized corona.  The data to date,
however, have been too sparse to firmly detail the properties of the
highly-ionized material in the LMC.

The \ovi\ absorption in this coronal scenario traces material at
relatively large distances from the disk that has been heated and
ejected from the disk by some form of feedback between massive stars
and the ISM.  Most calculations suggest this requires the collective
actions of multiple, correlated supernovae (particularly to raise the
material far above the plane).  There are several possible
microphysical origins for the \ovi\ within an extended corona about
the LMC.  Highly-ionized halo gas may trace material cooling from
higher temperatures through the transition temperatures (a
few$\times10^5$ K) at which \ovi\ is abundant or material created in
the interactions between cooler (e.g., $10^4$ K) and hotter (e.g.,
$>10^6$ K) gases (e.g., in conductive interfaces or turbulent mixing
layers).  These two broad types of models have significantly different
predictions for the \ovi\ columns through a hot corona.  As such, the
interpretation of the observed \ovi\ column density depends upon which
microphysical situation is most applicable.  We discuss the details of
each of these models below.

\subsubsection{Radiative Cooling Models}

Edgar \& Chevalier (1986) have calculated the column densities of
highly-ionized species for gas cooling from temperatures $T\ga 10^6$
K.  They show that for cases where the cooling is dominated by metal
line emission, the expected column density of \ovi\ and the other
high-ions is independent of metallicity.  The expected column density
of a metal ion, $Z^i$, in a column of cooling material is $N(Z^i)
\propto A_Z \dot{N} t_{\rm cool}$, where $A_Z$ is the abundance of the
element $Z$ with respect to hydrogen, $\dot{N}$ is the flux of cooling
material (in ionized hydrogen atoms \column\ s$^{-1}$), and $t_{\rm
cool}$ is the cooling time.  The cooling time is inversely
proportional to the abundance of metals, making $N(Z^i)$ independent
of $A_Z$.  This is important for the LMC because the abundance of
oxygen is $A_{\rm O} \approx 2.2\times10^{-4}$ (Russell \& Dopita
1992; see also Appendix of Welty et al. 1999), a factor of $\sim2.5$
below the recently updated solar value of $A_{\rm O} \approx
5.4\times10^{-4}$ (Holweger 2001).

Because $t_{\rm cool} \propto n_{\rm H^+}{}^{-1}$, where $n_{\rm H^+}$
is the initial ionized hydrogen density of the cooling material, the
column density of highly-ionized metals in the radiatively-cooling
flows discussed by Edgar \& Chevalier (1986) is proportional to
$\dot{N}/n_{\rm H^+}$, which, by conservation of mass is simply a
velocity, $v$.  The Edgar \& Chevalier calculations predict a column
density of $N(\mbox{\ovi})
\sim 4\times10^{14} \, (v/\mbox{100 \kms})$.  This value is sensitive
to the peak temperature from which the gas cools, which is assumed to
be $1\times10^6$ K.  If the peak temperature is $3\times10^6$ K, the
column density is increased by a factor of $\sim2$.  Using the mean
value of $N(\mbox{\ovi})$ from Table \ref{tab:statistics}, the
expected velocity associated with an Edgar \& Chevalier-like flow is
in the range $v \sim 30$ to 60 \kms.  The lower portion of this range
(which corresponds to higher initial cooling temperatures) is crudely
consistent with the observed velocity separation between the \ovi\ and
\feii\ absorption.

Since models of radiatively cooling gas predict the column density of
\ovi\ to be independent of metallicity, they provide a natural explanation
for the similarities between the values of $N_\perp$ in the Milky Way
and the LMC.  For the Edgar \& Chevalier models, so long as the ratio
$v \equiv \dot{N}/n_{\rm H^+}$ is similar in the two systems, the
column densities of the observed \ovi\ absorption should also be
similar.


In the context of a cooling flow like that described by Edgar \&
Chevalier (1986), the column density of \ovi\ is related to the mass
flow rate out of the disk, $\dot{M}$:
\begin{equation}
\dot{M} \approx  (\mu m_{\rm H})  \, n_{\rm H^+} \Omega
		\left( \frac{\dot{N}}{n_{\rm H^+}} \right).
\end{equation}
In this expression $\mu m_{\rm H}$ is the mean mass per atom, and
$\Omega$ is the surface area of the flow region, which we take to be
the disk of the LMC assuming a diameter of $\sim7.3$ kpc, the diameter
of the \HI\ disk (Kim et al. 1998).  For gas cooling under isobaric
conditions from $1\times10^6$ K, Edgar \& Chevalier find
$\dot{N}/n_{\rm H^+} \sim 2.5\times10^6 \, [N_\perp
(\mbox{\ovi})/10^{14} \ {\rm cm^{-2}}]$.  These values yield an
estimate for the mass-flow rate from the near side of the plane of the
LMC:
\begin{equation}
\dot{M} \sim 1 \, \left( \frac{n_{\rm H^+}}{10^{-2} \ 
	{\rm cm^{-3}}} \right) {\rm \ M_\odot \ yr^{-1}}.
\end{equation}
The initial density of the cooling gas, $n_{\rm H^+}$, is an unknown.
Points et al. (2001) estimate electron densities of the X-ray emitting
gas in the range $n_e \ga (0.3-2)\times10^{-2}$
\percc\ in their study of the X-ray emission from supergiant shells 
and diffuse gas in the LMC.  Dunne et al. (2001) estimate electron
densities of X-ray emitting gas in LMC superbubbles in the range $n_e
\ga (1-10)\times10^{-2}$ \percc.  These estimates depend upon the 
estimated volume occupied by the X-ray emitting material, and a
filling factor of unity for the material is assumed (the predicted
densities increase as the assumed filling factor decreases).  The
\ovi\ column densities themselves can also be used to crudely estimate
density limits for the hot ionized material.  If we assume that the
pathlength through LMC material probed by our observations is no
greater than the 7.3 kpc diameter of the LMC assumed above, the
average density of \ovi\ ions is $\langle n_{\rm O \, VI} \rangle
\equiv N(\mbox{\ovi})/d \ga 10^{-8}$ \percc.  This corresponds to
average ionized hydrogen densities of $\langle n_{\rm H} \rangle \ga
2\times10^{-4}$ \percc\ assuming the fraction of oxygen in the form of
\ovi\ is at its maximum expected value of $\sim0.2$ (Sutherland \& 
Dopita 1992).  The true density of ionized hydrogen in the \ovi
-bearing gas will likely be substantially larger since the fraction of
space occupied by this material is likely to be small.  We note,
however, that the densities suggested by this crude treatment of the
\ovi\ column densities and the X-ray analyses (which are biased towards 
high density material that is more likely to be in the plane of the
LMC) probably bound the physical ionized hydrogen density in the \ovi
-bearing gas.

The estimated mass-circulation rate from the LMC is a factor of $\sim
10$ lower than the values typically estimated for the Milky Way
assuming the same initial densities.  The main cause of this
difference is the much different surface areas assumed for the disks
of the two galaxies.  The mass-flow rate per unit area of the disk for
the LMC is
\begin{equation}
\dot{M} \sim 2\times10^{-2} \, 
	\left( \frac{n_{\rm H^+}}{10^{-2} \ {\rm cm^{-3}}} \right) {\rm
	\ M_\odot \ yr^{-1} \ kpc^{-2}},
\end{equation}
which should be closer to the Milky Way value for the same assumed
densities.

\subsubsection{Interface Models}

Models in which the \ovi\ ions occur in conductive interfaces or
turbulent mixing layers (TMLs) make different predictions for the
behavior of \ovi\ column density with abundance.  In the TML models,
for example, the column density of \ovi\ is related both to the
specific properties of the mixing (e.g., the velocity shear and
fraction of hotter material mixed into the turbulent layer) and the
abundance of the oxygen in the material (Slavin et al.  1993).
Therefore, in these models, the similarity in the Galactic and LMC
values of $N_\perp$ may simply be a coincidence dictated by the lower
abundance of the latter modified by a greater amount of transition
temperature material in the LMC compared with the Milky Way.

Increased amounts of gas in the \ovi -producing temperature range may
be related to different energy input into the ISM by massive stars
(and their corresponding supernovae), or to differences in the
interaction between hot and cool gas between the two galaxies
(determined by unknown mechanisms which may include the metallicity or
differences in the magnetic fields).  If we assume that the physics of
the mixing between the galaxies is similar, the increased quantities
of hot gas in the interface models requires a higher energy input into
the ISM in the LMC compared with the Milky Way.

A larger local energy input from massive stars in the LMC has some
observational support.  Tumlinson et al. (2001) have studied the
excitation and photodissociation of \htwo\ in the LMC.  Their models
predict that the near-ultraviolet (NUV) flux per unit area of the LMC
is a factor of $\sim10$ higher than in the solar neighborhood of the
Milky Way.  If the NUV radiation field in the LMC is taken to be an
indicator of the current density of massive stars, then this would
argue for a correspondingly larger kinetic energy input into the ISM
within the LMC through winds and supernovae.

Measures of the current star formation rate (SFR) and its surface
density (which is more closely related to the energy input per unit
area of the disk) also suggest the LMC should have a greater energy
input into the ISM than the Milky Way.  The SFR in the Milky Way is of
order 2 to 5 \msun\ yr$^{-1}$ (McKee \& Williams 1997; McKee 1989;
Mezger 1987), with average SFR per unit area of
$\sim(2-8)\times10^{-3}$ \msun\ yr$^{-1}$ kpc$^{-2}$, depending on the
adopted SFR and area over which the average is determined.  For the
LMC, the SFR is estimated to lie between 0.25 and 1 \msun\ yr$^{-1}$
(Kennicutt et al. 1995 and Klein et al. 1989, respectively).  The
former measurement is derived from \halpha\ observations while the
latter is from the thermal component of radio continuum measurements.
Given the probable effects of dust on the former, the higher value is
somewhat preferred.  Kim et al. (1998) determine that the bulk of the
\HI\ column of the LMC (which should correspond to the bulk of the
star formation) occurs within a disk of diameter $\sim7.3$ kpc.  This
implies a current SFR per unit area of $(6-24)\times10^{-3}$ \msun\
yr$^{-1}$ kpc$^{-2}$ (with the higher value preferred).  This also
suggests a larger energy input per unit area in the LMC than in the
local Milky Way, which would presumably lead to larger amounts of
highly-ionized gas.

Distinguishing between the radiatively-cooling outflow and hot/cool
gas interaction models as the origins of the observed
\ovi\ absorption will shed light on the structure and physics 
of a putative corona about the LMC and about the dependence of the
feedback efficiency in galaxies on metallicity and stellar energy
input.  However, the current observations give little leverage for
rejecting either class of models.  The ratios of highly-ionized atoms
in the gas can discriminate between pure radiatively cooling and
TML-type models (see Spitzer 1996).  Currently there are very few
reliable observations of the other high-ion transitions (e.g., of
\civ, \ion{Si}{4}, or \ion{N}{5}) in the LMC.  Bomans et al. (1996) 
have presented \hst\ observations of \civ\ towards two stars projected
within LMC 4, while Wakker et al. (1998) have presented \hst\
observations of \civ\ absorption seen towards five stars within the
``field.''  None of these objects are included in the current sample
of \fuse\ observations.

\subsection{Kinematics of the LMC Corona}

The kinematic profiles of \feii\ and \ovi\ along the sight lines in
this work suggest that the \ovi\ is kinematically decoupled from the
thin interstellar disk of the LMC.  The \ovi\ is, on average, observed
to be centered at velocities $\sim-30$ \kms\ from the \feii\
centroids.  There are two straightforward explanations for this
kinematic offset of the \ovi\ to more negative velocities: (1)
rotation of a thickened disk or halo; and (2) outflow of
highly-ionized material from the thin disk.  Both of these
explanations are compatible with our interpretation of the observed
\ovi\ properties of the LMC, i.e., that the \ovi\ is indeed tracing a
widely-distributed thick disk or halo of hot material.

A thickened disk of hot material that roughly rotates with the
underlying thin interstellar disk of the LMC could produce the
observed kinematic differences between the \ovi\ and the \feii.  The
ray from the Earth to a star in the disk of the LMC intersects the
thin disk (traced by \feii\ absorption) at essentially only the
location of the star; however, because it is vertically-extended, the
hot material (traced by \ovi\ absorption) is sampled over a greater
range of radial positions within the galaxy.  Under the assumption of
approximate corotation of disk and halo gas, this implies the \ovi\
absorption will be observed over a greater range of velocities.  In
particular, assuming the \HI\ velocity field (Kim et al. 1998) traces
the rotation of the thin interstellar disk, the \ovi\ is expected to
occur at velocities generally lower than the \feii\ in such a
scenario.  The maximum velocity of the \ovi\ should be approximately
equal to that expected of the thin disk at the position of the star,
i.e., equal to the central velocity of the \feii\ absorption.  This is
consistent with the velocity structure seen in Figure \ref{fig:phases}
(with a few exceptions).

A low-velocity outflow of highly-ionized material from the disk of the
LMC could also explain the observed kinematic differences between the
\ovi\ and \feii\ absorption profiles.  If we assume that the outflow is 
purely perpendicular to the plane of the LMC, then the outflow
velocity is $v_\perp \equiv \Delta v \sec i$, where $\Delta v$ is the
observed outflow velocity and $i$ is the inclination of the LMC from
the plane of the sky.  Our observations then imply $\langle v_\perp
\rangle \sim 40$ \kms.  The outflow of material from the disk is
predicted in all models for the production of an interstellar galactic
corona or thick disk.  However, it is not clear whether one expects to
observe \ovi\ associated with the outflow or the subsequent infall of
material in various models.  Models that produce \ovi\ in radiatively
cooling material require that the material ejected from the disk cool
sufficiently from the very high temperatures at which the gas
originates to temperatures where \ovi\ is expected to be abundant.
The observation of \ovi\ in the outflowing material of a fountain flow
would imply that the cooling time of the material was less than or of
the order the sound crossing time of the halo (Houck \& Bregman 1990).
If the cooling time is sufficiently long (e.g., for low densities or
very high initial temperatures), cooling through the temperatures at
which \ovi\ is abundant will not occur until the material ejected from
the disk reaches its highest point and begins to infall.

For all of the sight lines probed by our observations there is \ovi\
absorption at velocities that could imply the outflow of
highly-ionized (\ovi -bearing) gas from the thin disk of the LMC.
However, in only two cases (towards \sk68\dg80 and \sk70\dg91) is
there \ovi\ absorption at velocities that might imply the infall of
highly-ionized material.  If a large-scale circulation of material
through a corona about the LMC is occurring the majority of the gas
must cool in the outflow phase of the circulation.  We note that the
velocities of the observed \ovi\ absorption are low enough relative to
the disk of the LMC that the material will not escape the galaxy
altogether.


\section{Summary}
\label{sec:summary}

We have presented observations of 12 early-type stars within the Large
Magellanic Cloud obtained with the \fuse\ observatory.  These
observations reveal strong absorption from interstellar \ovi\ within
the LMC.  The main conclusions of our work are as follows.

\begin{enumerate}

\item Strong interstellar \ovi\ at LMC velocities is present along 
all of the sight lines studied in this work.  The observed stars probe
a range of interstellar environments from stars within super-giant
shells or superbubbles (e.g., \sk71\dg45, \sk68\dg80) to those in the
field (e.g., \sk67\dg20, \sk67\dg05).  A wide range of kinematic
profiles for \ovi\ within the LMC is also observed.

\item The column density of interstellar \ovi\ at LMC velocities ranges 
from $\log N(\mbox{\ovi}) = 13.9$ to 14.6 in atoms \column, with a
mean of $\log \langle N(\mbox{\ovi}) \rangle = 14.37$ (with a
dispersion of $\sim40\%$).  The average column density projected
perpendicular to the disk of the LMC is $\log \langle
N_\perp(\mbox{\ovi}) \rangle = 14.29$, which is identical to the value
derived from observations of the Milky Way halo even though the
metallicity of the LMC is $\sim2.5$ times lower than that of the Milky
Way.  The dispersions of the individual $N_\perp$ measurements about
the mean for each galaxy are also indistinguishable.

\item While our observations probe several sight lines projected onto 
superbubbles or other large-scale structures, any enhancements in the
\ovi\ column densities over those observed for field stars are minor 
($\la 50\% - 100\%$).  The column density variations over the smallest
scales probed by our observations ($\sim450-510$ pc) are the same as
those seen between sight lines towards and away from large-scale
structures.  Thus there is little evidence for enhanced \ovi\ column
densities towards superbubbles or other large-scale structures.

\item The average velocities of the  LMC \ovi\ absorption are 
different than those of the \feii\ absorption, which traces
low-ionization material associated with the disk of the LMC.  The
\ovi\ velocities are on average shifted by $\sim -30$ \kms\ from
the peak low-ionization absorption.  This systematic displacement of
the \ovi\ to lower velocities may be the signature of a low-velocity
outflow from the disk or of a rotating thickened disk.

\item  Our measurements are not able to test for the existence of high 
negative velocity \ovi\ in the LMC because of the overlap of Milky
Way, IVC, HVC, and LMC absorption.  The observations of all of the
sight lines considered here are consistent with the presence of \ovi\
outflowing from the LMC disk at relatively low velocities ($\la 75$
\kms).  Material at  positive velocities relative to the LMC disk,
which could trace infalling material, is only allowed along the sight
lines towards \sk68\dg80 and possibly \sk70\dg91.  The former sight
line shows gas which may be infalling at velocities $\sim70$ \kms\
relative to the LMC disk.

\item The velocity dispersion of the \ovi\ absorption is much greater
than that of \feii\ and other low-ionization species seen along the
same lines of sight. The former is difficult to measure due to overlap
between Milky Way and LMC absorption along many sight lines, but where
measurable gives Gaussian dispersions of $\sigma \sim 30$ to 50 \kms.
These breadths are much broader than the expected thermal broadening
($\sigma_{thermal} \sim 12$ \kms\ for $T\sim3\times10^5$ K),
suggesting the presence of several absorbing components within the LMC
\ovi\ profiles.

\item We discuss the observed properties of interstellar LMC \ovi\ in
the context of a galactic ``halo'' or ``corona'' similar to that of
the Milky Way.  The striking similarity in the \ovi\ columns and
variations and in the observed velocity dispersions between the LMC
absorption and that observed in the halo of the Milky Way motivate
such a model.  The observations are consistent with models of
radiatively-cooling galactic fountain flows, although turbulent mixing
layers and other types of interfaces are not ruled out.  The
radiatively-cooling fountain scenario is attractive since the
predicted column density of \ovi\ is not dependent on the metallicity
of the cooling material.

\item If the \ovi\ is produced in a galactic fountain flow, then the 
mass flow from one side of the disk of the LMC is estimated to be
$\dot{M} \sim 1$ \msun\ yr$^{-1}$, or alternatively, the mass flow
rate per unit area of the disk is estimated to be $\dot{M}/\Omega \sim
2\times10^{-2}$ \msun\ yr$^{-1}$ kpc$^{-2}$.  The cooling of this
ejected material occurs before the gas begins to return to the disk.
The average density of ionized hydrogen associated with the \ovi
-bearing gas likely exceeds $n_{\rm H^+} \ga 2\times10^{-4}$ \percc;
studies of X-ray emission in the LMC suggest $n_{\rm H^+} \sim
10^{-2}$ \percc, though this may sample denser gas than the \ovi
-bearing material.  Better estimates of the physical state of the gas
could be made with \ovi\ emission line measurements towards the sight
lines probed by our absorption line measurements.

\end{enumerate}

\acknowledgements

We thank J. Gaustad and collaborators for the use of their \halpha\
image of the LMC.  This image is from the Southern H-Alpha Sky Survey
Atlas (SHASSA), which is supported by the National Science Foundation.
We also thank Y.-H. Chu for help with the \rosat\ X-ray mosaic.  This
work is based on data obtained for the Guaranteed Time Team by the
NASA-CNES-CSA FUSE mission operated by the Johns Hopkins
University. Financial support to U.S. participants has been provided
by NASA contract NAS5-32985.  JCH and KRS also recognize support from
NASA Long Term Space Astrophysics grant NAG5-3485 through the Johns
Hopkins University.


%
%
%



\newcommand{\fe}[1]{\ensuremath{\times10^{#1}}}

\begin{deluxetable}{llllclccc}
\setlength{\tabcolsep}{0.06in} 
\tablenum{1}
\tablecolumns{10}
\tablewidth{0pt}
\tablecaption{Sightline and Stellar Properties for {\em FUSE}
        LMC Targets \label{tab:targets}}
\tablehead{
\colhead{Star} & \colhead{Alt. Name} &
\colhead{$\alpha$ (J2000)} & \colhead{$\delta$ (J2000)} &
\colhead{V [mag]} &
\colhead{Sp. Type} & \colhead{Ref.\tablenotemark{a}} &
\colhead{$v_\infty$ [km/s]\tablenotemark{b}} 
& \colhead{Ref.\tablenotemark{c}}
}
\startdata
Sk--67\dg05\tablenotemark{d}
 & HD 268605 & $04^h50^m18\fs9$ & --$67^\circ 39\arcmin 38\farcs2$ &
        11.34 & O9.7 Ib & 1 & 1665 & 1 \\
Sk--67\dg20 & HD 32109 & 04 55 31.5 & --67 30 01.0 &
        13.87 & WN4b & 2 & 2900 & 2 \\ 
Sk--66\dg51 & HD 33133 & 05 03 10.2 & --66 40 54.0 &
        12.69 & WN8h & 2 & ~850 & 3 \\ 
Sk--67\dg69  & \nodata & 05 14 20.2 & --67 08 03.5 &
        13.09 & O4 III(f) & 3 & 2600 & 4 \\ 
Sk--68\dg80 & HD 36521 & 05 26 30.4 & --68 50 26.6 &
        12.40 & WC4+OB & 4 & \nodata & \nodata \\ 
Sk--70\dg91 & LH 62-1 & 05 27 33.7 & --70 36 48.3 &
        12.78 & O6.5 V & 5 & 3000 & 5 \\ 
Sk--66\dg100 & \nodata & 05 27 35.6 & --66 55 15.0 &
        13.26 & O6 II(f) & 6 & 2075 &  6 \\ 
Sk--67\dg144 & HD 37026 & 05 30 12.2 & --67 26 08.4 &
        14.30 & WC4 & 4 & 2600 & 7 \\
Sk--71\dg45 & HD 269676 & 05 31 15.5 & --71 04 08.9 &
        11.47 & O4-5 III(f) & 1 & 2500 & 5 \\ 
Sk--69\dg191 & HD 37680 & 05 34 19.4 & --69 45 10.0 &
        13.35 & WC4 & 4 & 2800 &  7 \\ 
Sk--67\dg211 & HD 269810 & 05 35 13.9 & --67 33 27.0 &
        12.26 & O2 III(f$^\ast$) & 7 & 3750 & 8 \\ 
Sk--66\dg172 & \nodata & 05 37 05.6 & --66 21 35.7 &
        13.13 & O2 III(f$^\ast$)+OB & 7 & 3250 & 8 \\ 
\enddata
\tablenotetext{a}{Reference for the adopted spectral classification:
        (1) Walborn 1977;
        (2) Smith, Shara, \& Moffat 1996;
        (3) Garmany \& Walborn 1987;
        (4) Smith, Shara, \& Moffat 1990;
        (5) Conti, Garmany, \& Massey 1986;
        (6) Walborn et al. 1995;
        (7) Walborn et al. 2001.}
\tablenotetext{b}{Stellar wind terminal velocity.  The terminal 
        velocities for the WR stars often have larger uncertainties
        than those for the O stars.}
\tablenotetext{c}{Reference for the adopted wind terminal velocities:
        (1) Patriarchi \& Perinotto 1992;
        (2) Koesterke et al. 1991;
        (3) Crowther \& Smith 1997;
        (4) Leitherer 1988; 
        (5) Massa et al. 2001;
        (6) Prinja \& Crowther 1998;
        (7) Gr{\"a}fener et al. 1998;
        (8) Walborn et al. 1995. }
\tablenotetext{d}{The Sk--67\dg05 data used here are from
        the early-release observations of Friedman et al. 2000.}
\end{deluxetable}

\begin{deluxetable}{lllccc}
\tablenum{2}
\tablecolumns{6}
\tablewidth{0pt}
\tablecaption{Log of {\em FUSE} Observations \label{tab:log}}
\tablehead{
\colhead{Star} & 
\colhead{FUSE} & 
\colhead{Start} &
\colhead{No. of} & 
\colhead{Exp. Time} & \colhead{S/N} \\
\colhead{} & \colhead{ID\tablenotemark{a}} & 
\colhead{Date\tablenotemark{b}} &
\colhead{Exp.\tablenotemark{c}} & 
\colhead{[ksec]} & \colhead{($\lambda1031$)\tablenotemark{d}}
}
\startdata
Sk--67\dg05 & \multicolumn{1}{c}{\tablenotemark{e}} & 
	\multicolumn{1}{c}{\tablenotemark{e}} & 
	\multicolumn{1}{c}{\tablenotemark{e}} & 
	$33.0$ & $>30$ \\
%
Sk--67\dg20  & P11744 & 10/12/2000 & 6 & 16.3 & 20 \\
Sk--66\dg51  & P11745 & 09/30/2000 & 4 &  4.6 & 18 \\
Sk--67\dg69  & P11717 & 12/20/1999 & 4 &  7.8 & 15 \\
Sk--68\dg80  & P10314 & 12/17/1999 & 4 &  9.7 & 24 \\
Sk--70\dg91  & P11725 & 10/05/2000 & 1 &  5.5 & 18 \\
Sk--66\dg100 & P11723 & 12/20/1999 & 2 &  7.1 & 17 \\
Sk--67\dg144 & P11750 & 02/12/2000 & 3 &  8.1 & 10 \\
Sk--71\dg45  & P10315 & 10/02/2000 & 4 & 18.9 & 25 \\
Sk--69\dg191 & P11751 & 02/12/2000 & 3 &  7.0 & 11 \\
Sk--67\dg211 & P11716 & 12/20/1999 & 5 &  8.2 & 21 \\
Sk--66\dg172 & P11722 & 09/26/2000 & 1 &  3.6 & 13 \\
\enddata
\tablenotetext{a}{Archival rootname of target for FUSE PI team 
	observations.  All data were processed with versions 1.8.6 or
	1.8.7 of the {\tt CALFUSE} pipeline.}
\tablenotetext{b}{Start date of the observations.  For the data 
	presented here, the data were all collected within a few days
	of the first exposures.}
\tablenotetext{c}{Number of individual exposures.}
\tablenotetext{d}{Approximate signal-to-noise ratio per 
	20 km s$^{-1}$ resolution element near the strong \ion{O}{6}
	transition at 1031.926 \AA\ in the LiF1 channel.}
\tablenotetext{e}{The observations of this star have been described by 
	Friedman et al. 2000.  These data required special
	processing since the star was stepped across the $30\arcsec
	\times 30 \arcsec$ LWRS aperture during the integrations.  The
	data were acquired at various times between 08/20/1999 and
	10/19/1999.}
\end{deluxetable}

\begin{deluxetable}{cccl}
\tablenum{3}
\tablecolumns{4}
\tablewidth{0pt}
\tablecaption{Molecular Hydrogen Lines\tablenotemark{a} \label{tab:htwo}}
\tablehead{
\colhead{$\lambda$ [\AA]} & \colhead{ID} & 
\colhead{$\log \lambda f$} & \colhead{Notes}
}
\startdata
\cutinhead{$J=3$}
{\bf 1031.191} & {\bf (6-0) P(3)} & {\bf 1.059} & 
	{\bf Contaminating line} \\
1006.411 & (8-0) R(3) & 1.199 & \\
1019.500 & (7-0) P(3) & 1.029 & \\
1028.985 & (6-0) R(3) & 1.253 & Blended: Ly$\beta$ \\
1043.502 & (5-0) P(3) & 1.051 & Blended: MW R(4) \\
1053.976 & (4-0) R(3) & 1.149 & \\
1056.472 & (4-0) P(3) & 1.004 & Blended: MW R(4) \\
1067.479 & (3-0) R(3) & 1.030 & \\
\cutinhead{$J=4$}
{\bf 1032.349} & {\bf (6-0) R(4)} & {\bf 1.247} & 
	{\bf Contaminating line} \\
 999.268 & (9-0) R(4) & 1.219 & \\
1023.434 & (7-0) P(4) & 1.031 & \\
1035.181 & (6-0) P(4) & 1.067 & \\
1044.542 & (5-0) R(4) & 1.209 & Blended: LMC P(3) \\
1057.380 & (4-0) R(4) & 1.136 & Blended: LMC P(3) \\
1060.581 & (4-0) P(4) & 1.017 & \\
\enddata
\tablenotetext{a}{Molecular hydrogen lines used for determining 
   	contamination of interstellar \ion{O}{6} profiles.  The (6-0)
   	P(3) line at LMC velocities and (6-0) R(4) line at Milky Way
   	(MW) velocities contaminate the \ion{O}{6} $\lambda1031.926$
   	line when present. Lines marked as being blended refer to the
   	LMC and Milky Way components for the $J=3$ and 4 transitions,
   	respectively.  All of the listed lines are Lyman transitions.}
\end{deluxetable}

\begin{deluxetable}{rllll}
\tablenum{4}
\tablecolumns{7}
\tablewidth{0pt}
\tablecaption{Equivalent Widths and Column Densities of 
	Interstellar \ion{O}{6} in the LMC \label{tab:columns}}
\tablehead{
\colhead{ID} & \colhead{Star} & 
\colhead{$W_\lambda$ [m\AA]\tablenotemark{a}} &
\colhead{$\log N(\mbox{\ion{O}{6}})$\tablenotemark{b}} &
\colhead{$v_-,v_+$\tablenotemark{c}}
}
\startdata
 1 &  Sk--67\dg05\tablenotemark{d} & $  87\pm 5$ & $13.89^{+0.07}_{-0.06}$ & $+180,+325$  \\
 2 &  Sk--67\dg20 & $ 180\pm10$ & $14.26^{+0.09}_{-0.10}$ & $+175,+335$  \\
 3 &  Sk--66\dg51 & $ 199\pm21$ & $14.31^{+0.07}_{-0.07}$ & $+180,+365$  \\
 4 &  Sk--67\dg69 & $ 249\pm11$ & $14.48^{+0.03}_{-0.03}$ & $+160,+345$  \\
 5 &  Sk--68\dg80 & $ 331\pm 7$ & $14.61^{+0.03}_{-0.04}$ & $+140,+330$  \\
 6 &  Sk--70\dg91 & $ 297\pm 9$ & $14.55^{+0.07}_{-0.08}$ & $+175,+375$  \\
 7 & Sk--66\dg100 & $ 164\pm15$ & $14.26^{+0.06}_{-0.05}$ & $+175,+315$  \\
 8 & Sk--67\dg144 & $ 220\pm16$ & $14.41^{+0.07}_{-0.06}$ & $+175,+335$  \\
 9 &  Sk--71\dg45 & $ 223\pm 5$ & $14.38^{+0.05}_{-0.06}$ & $+185,+350$  \\
10 & Sk--69\dg191 & $ 205\pm18$ & $14.38^{+0.09}_{-0.10}$ & $+175,+330$  \\
11 & Sk--67\dg211 & $ 164\pm 6$ & $14.21^{+0.04}_{-0.04}$ & $+185,+375$  \\
12 & Sk--66\dg172 & $ 191\pm14$ & $14.31^{+0.05}_{-0.06}$ & $+195,+365$  \\
\enddata
\tablenotetext{a}{Equivalent widths for the LMC material along the 
	observed sight lines with $1\sigma$ error estimates.  The error
	estimates include the effects of shifting the lower velocity
	integration limit by $\pm20$ km s$^{-1}$.}
\tablenotetext{b}{\ion{O}{6} column densities for LMC material 
	along the observed sight lines with $1\sigma$ error estimates.
	In all cases these column densities have been derived using
	observations of the 1031.926 \AA\ transition assuming no
	unresolved saturation is present.  We adopt an $f$-value of $f
	= 0.1325$ from the theoretical calculations of Yan, Tambasco,
	\& Drake 1998.}
\tablenotetext{c}{Velocity range over which the LMC profile was integrated.}
\tablenotetext{d}{The \ion{O}{6} column densities towards 
	Sk--67$^\circ \,$05 are taken from Table 2 of Friedman et
	al. 2000 assuming their ``upper'' continuum placement.}
\end{deluxetable}

\begin{deluxetable}{lrr}
\tablenum{5}
\tablecolumns{3}
\tablewidth{0pt}
\tablecaption{Statistical Properties of Interstellar \ion{O}{6} in the LMC \tablenotemark{a} 
		\label{tab:statistics}}
\tablehead{
\colhead{Quantity} & \colhead{Linear} & \colhead{Logarithmic}

}
\startdata
\cutinhead{Full Sample} 
Mean 		&  $2.34\fe{14}$ 	& 14.37 \\
Weighted Mean 	&  $1.80\fe{14}$	& 14.25 \\
Std. Deviation	&  $\pm0.89\fe{14}$	&  $^{+0.14}_{-0.21}$ \\
Median		&  $2.37\fe{14}$	& 14.38 \\
\cutinhead{New Observations\tablenotemark{b}}
Mean 		&  $2.49\fe{14}$ 	& 14.40 \\
Weighted Mean 	&  $2.22\fe{14}$	& 14.35 \\
Std. Deviation	&  $\pm0.78\fe{14}$	&  $^{+0.12}_{-0.16}$ \\
Median		&  $2.37\fe{14}$	& 14.38 \\
\enddata
\tablenotetext{a}{All quantities quoted were derived from the linear column densities. } 
\tablenotetext{b}{Statistics for the sight lines excluding that towards Sk--67$^\circ \,$05.
	This star, described in detail by Friedman et al. 2000, has
	continuum placement uncertainties that likely exceed those of
	the rest of our sample.}
\end{deluxetable}

%

\begin{deluxetable}{clll}
\tablenum{6}
\tablecolumns{4}
\tablewidth{0pt}
\tablecaption{Interstellar Environments of the Background Probes\label{tab:environs}}
\tablehead{
\colhead{ID\tablenotemark{a}} & 
\colhead{Star} & 
\colhead{Nebulosity\tablenotemark{b}} &
\colhead{Description} 
}
\startdata
1 & Sk--67\dg05 & N3/DEM7 & Diffuse, faint \ion{H}{2} region  \\                   
2 & Sk--67\dg20  & \nodata & Field star  \\                                        
3 & Sk--66\dg51  & DEM56 & Faint \ion{H}{2} region \\                              
4 & Sk--67\dg69  & DEM107 & Faint \ion{H}{2} region on periphery of KDSB GS 44 \\  
5 & Sk--68\dg80 & N144/DEM199 & Superbubble projected onto LMC 3 \\                
6 & Sk--70\dg91  & N204/DEM208 & Superbubble \\                                    
7 & Sk--66\dg100 & \nodata & Interior of LMC 4 \\                                  
8 & Sk--67\dg144 & \nodata & Periphery of LMC 4 \\                                 
9 & Sk--71\dg45 & N206/DEM221/KDSB GS70 & Superbubble \\                           
10 & Sk--69\dg191 & N154/DEM246/KDSB GS72 & Superbubble \\                         
11 & Sk--67\dg211 & N59A/DEM241 & Bright \ion{H}{2} region \\                      
12 & Sk--66\dg172 & N64B/DEM252 & Bright \ion{H}{2} region on periphery of LMC 4 \\
\enddata
\tablenotetext{a}{Identifications corresponding to the labels in 
	Figures 5, 6, and 7.}
\tablenotetext{b}{The designations N and DEM refer to entries in the 
	catalogs of Henize 1956 and Davies, Elliott, \& Meaburn 1976,
	respectively.  The designation KDSB GS refers to an entry in
	the Kim et al. 1999 catalog of ``giant shells'' identified in
	the Kim et al. 1998 \ion{H}{1} mosaic of the LMC.}
\end{deluxetable}

\clearpage


\end{document}